\documentclass[twocolumn,showpacs,preprintnumbers,amsmath,amssymb,pre]{revtex4}

\usepackage{graphicx}
\usepackage{dcolumn}
\usepackage{bm}
\usepackage{color}

\begin{document}

\title{Critical behavior of self-assembled rigid rods on two-dimensional lattices: Bethe-Peierls approximation and Monte Carlo simulations}

\author{L. G. L\'opez}
\author{D. H. Linares}
\author{A. J. Ramirez-Pastor}
\email{antorami@unsl.edu.ar} \affiliation{Departamento de
F\'{\i}sica, Instituto de F\'{\i}sica Aplicada, Universidad
Nacional de San Luis, CONICET, 5700 San Luis, Argentina}

\author{D. A. Stariolo}
\affiliation{Departamento de F\'{\i}sica, Universidade Federal do Rio Grande do Sul and
National Institute of Science and Technology for Complex Systems
CP 15051, 91501-970 Porto Alegre, RS, Brazil}

\author{S. A. Cannas}
\affiliation{Facultad de Matem\'atica, Astronom\'{\i}a y
F\'{\i}sica, Universidad Nacional de C\'ordoba and \\ Instituto de
F\'{\i}sica Enrique Gaviola  (IFEG-CONICET), Ciudad Universitaria,
5000 C\'ordoba, Argentina}

\date{\today}

\begin{abstract}

The critical behavior of adsorbed monomers that
reversibly polymerize into linear chains with restricted
orientations relative to the substrate has been studied.
In the model considered here, which is known as self-assembled rigid rods (SAARs)
model, the surface is represented by a two-dimensional lattice
and a continuous orientational transition occurs as a function of
temperature and coverage. The phase diagrams were
obtained for the square, triangular and honeycomb lattices by
means of Monte Carlo simulations and finite-size scaling analysis.
The numerical results were compared with Bethe-Peierls analytical
predictions about the orientational transition for the square and
triangular lattices. The analysis of the phase diagrams, along
with the behavior of the critical average rod lengths, showed that
the critical properties of the model do not depend on the
structure of the lattice at low temperatures (coverage),
revealing a one-dimensional behavior in this regime. Finally, the
universality class of the SAARs model, which has been subject of
controversy, has been revisited.

\end{abstract}


\maketitle

\section{INTRODUCTION}

Self-assembly has become a topic of increasing interest in recent
years. One reason for this interest is that it is central to
understanding structure formation in living systems \cite{Workum}.
As a consequence, a significant research effort has been devoted
to enhance our understanding of the theoretical basis of the
fundamental mechanisms governing self-assembly and the observables
required to characterize the interactions driving thermodynamic
self-assembly transitions \cite{Zhang,Palma}. More related to the
present work, several research groups reported on the assembly of
particles in linear chains
\cite{Lu,Scior,Ouyang,Tavares0,Sciortino}. Despite of the number
of contributions to this problem, the knowledge of how this
process works is still limited.

It is obvious that a complete analysis of the self-assembly
phenomenon is quite a difficult subject because of
the complexity of the involved microscopic mechanisms. For
this reason, the understanding of simple models with increasing
complexity might be a help and a guide to establish a
general framework for the study of this kind of systems, and
to stimulate the development of more sophisticated models
which can be able to reproduce concrete experimental situations.

In this context, an interesting model was introduced by Tavares et
al. \cite{Tavares}. The system in Ref. [\onlinecite{Tavares}]
consists of monomers with two attractive (sticky) poles that
polymerize reversibly into polydisperse chains and, at the same
time, undergo a continuous isotropic-nematic (IN) phase
transition. Using an approach in the spirit of the Zwanzig model
\cite{Zwanzig}, the authors studied the IN transition occurring in
the system. The obtained results revealed that nematic ordering
enhances bonding. In addition, the average rod length was
described quantitatively in both phases, while the location of the
ordering transition, which was found to be continuous, was
predicted semiquantitatively by the theory. Finally, Tavares et
al. assumed as working hypothesis that the nature of the IN
transition remains unchanged with respect to the case of
monodisperse rigid rods on square lattices, where the transition
is in the 2D Ising universality class
\cite{GHOSH,Matoz,Matoz1,Matoz2}.

From the seminal work by Tavares et al., a series of papers
exploring the self-assembled rigid rods (SARRs) model have been
published
\cite{Lopez1,Almarza1,Lopez4,Almarza3,Lopez2,Erratum,Almarza2,Lopez3,Lopez5}.
These studies can be separated in two groups: (i) those dealing
with the nature and universality of the phase transition occurring
in the system
\cite{Lopez1,Almarza1,Lopez4,Almarza3,Lopez2,Erratum,Almarza2} and
(ii) those dealing with the temperature-coverage phase diagram of
the system \cite{Lopez3,Lopez5}.

With respect to the first point, the universality class of the
model has been a subject of controversy. Thus, the criticality of
the SARRs model in the square  lattice was investigated in Ref.
[\onlinecite{Lopez1}] by means of canonical Monte Carlo (MC)
simulation and finite-size scaling (FSS) theory. The existence of
a continuous phase transition was confirmed.
In addition, the determination of the critical exponents, along with the behavior of
the Binder cumulants, revealed that the universality class of the IN transition
changes from 2D Ising-type for monodisperse RRs without
self-assembly to $q = 1$ Potts-type (random percolation) for
polydisperse SARRs. Later, a multicanonical MC method based on a
Wang-Landau sampling scheme was used by Almarza et al.
\cite{Almarza1} to reinvestigate the critical behavior of the
model studied in Refs. [\onlinecite{Tavares,Lopez1}]. Employing
the crossing point of the Binder cumulants and the value of the
critical exponent of the correlation length ($\nu$), it was
observed that the criticality of the SARRs model is in the 2D
Ising class, as in models of monodisperse RRs \cite{Matoz,Matoz2}.

The results in Refs. [\onlinecite{Lopez1,Almarza1}], along with
the recent study in Ref. [\onlinecite{Lopez4}], indicate that the
system under study represents an interesting case where the use of
different statistical ensembles (canonical or grand canonical)
leads to different and well-established universality classes
($q=1$ Potts-type or $q=2$ Potts-type, respectively). A similar
scheme was observed for triangular lattices, where canonical MC
simulations indicated that the IN transition of SARRs at
intermediate density belongs to the q=1 Potts universality class
\cite{Lopez2,Erratum}. In contrast with this result, a $q=3$
Potts-type universality was obtained by using grand canonical MC
simulations \cite{Almarza2}.

Among the studies of the second group, the temperature-coverage
phase diagram of SARRs on square lattices was calculated in Ref.
[\onlinecite{Lopez3}]. By using MC simulations, mean-field theory,
and a renormalization group (RSRG) approach, the critical line
which separates regions of isotropic and nematic stability was
obtained and characterized. The results showed that the theory
presented in Ref. [\onlinecite{Tavares}] overestimates the
critical temperature in all range of coverage. Small differences appear between simulation and
theoretical results for small values of $\theta$; however, the disagreement turns out to be
significantly large for larger $\theta$'s. On the other hand, RSRG reproduces qualitatively the shape of the critical
line, but systematically underestimates the critical temperature.
The main prediction of RSRG approach is that the critical properties of the whole line are associated to a unique
second-order fixed point, confirming the continuous nature of the
transition. Concerning the MF results, the theory predicts the
existence of a first-order transition line and a tricritical
point. This finding is in sharp contrast to that obtained by MC
simulations and RSRG approach.

More recently, the main adsorption properties of SARRs on square
and triangular lattices have been addressed \cite{Lopez5}. The
study demonstrated that the adsorption isotherms appear as
sensitive quantities to the IN phase transition, allowing to
reproduce the temperature-coverage phase diagram of the system for
square lattices, and to obtain a first determination of the phase
diagram for triangular lattices.

Following the line of Refs.
[\onlinecite{Lopez1,Almarza1,Lopez4,Almarza3,Lopez2,Erratum,Almarza2,Lopez3,Lopez5}],
the present paper deals with the two aspects above mentioned. On
one hand, the problem of the universality is revisited, clarifying
recent results obtained for triangular lattices (with conclusions
that can be extrapolated to the honeycomb lattice case). On the
other hand, the complete temperature-coverage phase diagrams were
obtained for the square, triangular and honeycomb lattices by
means of MC simulations and FSS analysis. The critical lines were
also calculated for the square and triangular lattices within the
Bethe-Peierls (BP) or quasi-chemical approach, as formulated in
the Cluster Variational Method. Comparisons with MF and RSRG data
indicate that BP represents a qualitative advance in the
analytical description of the phase diagram of SARRs.

The rest of the paper is organized as follows. In Sec. II we
describe the lattice-gas model. The simulation scheme and
computational results are given in Sec. III. In Sec. IV we present
the analytical approximations, and compare the
MC results with the theoretical calculations. In Sec. V we review
previous results on the universality class of the SAARs model in the
triangular and honeycomb lattices.
Finally, the general conclusions are drawn in Sec. VI.


\section{LATTICE-GAS MODEL}
\label{Lattice-Gas Model}

As in Refs. [\onlinecite{Tavares,Lopez1,Lopez2,Erratum,Lopez3,Lopez4,Lopez5,
Almarza1,Almarza2,Almarza3}], a system of self-assembled rods with
a discrete number of orientations in two dimensions is considered.
The substrate is represented by a square, triangular or honeycomb
lattice of $M = L \times L$ adsorption sites, with periodic boundary
conditions. $N$ particles are adsorbed on the substrate with $m$
possible orientations along the principal axes of the array, being
$m=2$ for square lattices and $m=3$ for triangular and honeycomb
lattices. The rods interact with nearest-neighbors (NN)
through anisotropic attractive interactions. Thus, a cluster or
uninterrupted sequence of bonded particles is a self-assembled rod.
Then, in the canonical ensemble the adsorbed phase is characterized
by the Hamiltonian

\begin{equation}
H = -w \sum_{\langle i,j \rangle} \Big \{ |\vec{r}_{ij} \cdot
\vec{\sigma}_i||\vec{r}_{ji} \cdot \vec{\sigma}_j| \ {\rm div} \ 1 \Big \}, \label{hc}
\end{equation}

\noindent where $\langle i,j \rangle$ indicates a sum over NN
sites; $w$ represents the NN lateral interaction between two
neighboring sites $i$ and $j$; the energy is lowered by an amount $|w|$
only if the NN monomers are aligned with each other and with the
intermolecular vector $\vec{r}_{ij}$; $\vec{\sigma}_j$ is the
occupation vector, with $\vec{\sigma}_j=0$ if the site $j$ is
empty and $\vec{\sigma}_j= \hat{x}_k$ if the site $j$ is occupied
by a particle with orientation parallel to $x_k$. $\hat{x}_k$ are
a set of unit vectors along the crystalline axes. In Eq.(\ref{hc}) ,
{\em div} represents an integer division, so the result inside \{
\} can be either 0 or 1 (i.e. the fractional part is discarded).
The integer division is redundant in the case of the Hamiltonian
for the square lattice, but it avoids additional lateral
interactions \cite{foot0} that promote the condensation of the
monomers in the triangular and honeycomb lattice \cite{Erratum},
restricting the attractive couplings only to those pairs of NN
monomers whose orientations are aligned with each other and with
the monomer-monomer lattice direction, in line with the model in
the square lattice.

The grand canonical Hamiltonian of the model is given by
\begin{equation}
H = -w \sum_{\langle i,j \rangle} \Big \{ |\vec{r}_{ij} \cdot
\vec{\sigma}_i||\vec{r}_{ji} \cdot \vec{\sigma}_j| \ {\rm div} \ 1 \Big \}
- \left(\mu- \epsilon_o \right) \sum_i |\vec{\sigma}_i|, \label{hgc}
\end{equation}
\noindent where $\epsilon_o$ is the adsorption energy of an adparticle
on a site and $\mu$ is the chemical potential. In the present work,
$\epsilon_o=0$ and $\mu$ is the only parameter that determine the strength
of the adsorption.

On the other hand, it is worth mentioning that, while the concept of linear rod
is trivial for square and triangular lattices, in a honeycomb lattice
the geometry does not allow for the existence of a linear array of monomers.
In this case, we call linear rod to a chain of adjacent monomers that
can be assembled in only three types of sequences, defining three directions
in similar way to the triangular lattice. For more details about the model
in the honeycomb lattice, see Ref. [\onlinecite{Lopez2}].

In the case of a square lattice the grand canonical Hamiltonian
(\ref{hgc}) can be exactly mapped into a spin-1 model
\cite{Lopez3}

\begin{eqnarray}
     H &=&  -\frac{w}{4} \sum_{<i,j>} \left[(S_i^2+S_i)(S_j^2+S_j)(\hat{x}_2 . \vec{r}_{ij})+\right. \nonumber \\
        & & \left.+  (S_i^2-S_i)(S_j^2-S_j)(\hat{x}_1.\vec{r}_{ij})\right]-\mu \sum_i S_i^2   \label{HS1}
\end{eqnarray}

\noindent where $S_i=0,\pm 1$ and $\hat{x}_1,\hat{x}_2$ are unit
vectors along the two orthogonal crystalline directions. $S_i\pm
1$ represent the vertical ($\vec{\sigma}_i= \hat{x}_2$) and the
horizontal ($\vec{\sigma}_i= \hat{x}_1$) orientations, while
$S_i=0$ represents the empty state.

In the case of a triangular lattice the model can be formulated in
terms of a diluted $q=3$ anisotropic Potts model. We associate to
each site of the lattice  a spin variable $\sigma_i=0,1,2,3$, such
that $\sigma_i=0$ represents the empty state and the states
$\sigma_i=1,2,3$ represent a bar oriented along the three unit
vectors $\hat{x}_k$ ($k=1,2,3$), where the angle between any pair
of them is $ 2\pi/3$ . Then the Hamiltonian can be written as

\begin{eqnarray}\label{Hpotts}
     H &=& -w \sum_{<i,j>} \sum_{\sigma=1}^3 \delta(\sigma_i,\sigma)\, \delta(\sigma_j,\sigma)\, \delta(\vec{r}_{ij}, \pm \hat{x}_\sigma) -\nonumber\\
     & & -\mu \sum_i \left[1-\delta(\sigma_i,0) \right]\label{Hpotts}
\end{eqnarray}

\noindent where $\delta(\sigma,\sigma')$ is the Kronecker delta
function. These alternative representations of the model are
useful for the analytical treatment.

\section{MC SIMULATIONS}

\subsection{MC method}

We have used a standard importance sampling MC method in the
canonical and grand canonical ensembles \cite{BINDER} and FSS
techniques \cite{PRIVMAN}. All calculations were carried out using
the parallel cluster BACO of Universidad Nacional de San Luis,
Argentina.

\subsubsection{Canonical MC simulations}

MC simulations in the canonical ensemble were used to obtain the
results presented in the subsection III.B. The procedure is as
follows. Starting with a random initial configuration (sites
occupied with concentration $\theta=N/M$ and particle axis
orientation chosen at random), successive configurations are
generated by attempting to move single particles (monomers). One
of the two (translation or rotation) moves is chosen at random. In
a translation move, an occupied site and an empty site are
randomly selected and their coordinates are established. Then, an
attempt is made to interchange its occupancy state with
probability given by the Metropolis rule \cite{Metropolis}: $P =
\min \left\{1,\exp\left( - \beta \Delta H \right) \right\}$, where
$\Delta H$ is the difference between the Hamiltonians of the final
and initial states and $\beta=1/k_BT$ (being $k_B$ the Boltzmann
constant). For a rotation move, the rotational state of the chosen
particle is changed with a probability determined by Metropolis
rule. A MC step (MCS) is achieved when $\theta \times M$ sites
have been tested to change its occupancy state. Typically, the
equilibrium state can be well reproduced after discarding the
first $5 \times 10^6$ MCS. Then, the next $6 \times 10^8$ MCS are
used to compute averages.

\subsubsection{Grand canonical MC simulations}

MC simulations in the grand canonical ensemble have been carried
out to understand the discrepancy between the results of
\cite{Lopez2} and \cite{Almarza2} about the universality class of
SAARs model in the triangular lattice (Sec. V). The procedure is
as follows. For a given pair of values of $T$ and $\mu$, an
initial configuration with $N$ monomers adsorbed at random
positions and orientations (on $M$ sites) is generated. Then an
adsorption-desorption process is started, where the lattice sites
are tested to change its occupancy state with probability given by
the Metropolis rule \cite{Metropolis}: $P = \min
\left\{1,\exp\left( - \beta \Delta H \right) \right\}$, where
$\Delta H$ is the difference between the Hamiltonians of the final
and initial states and $\beta=1/k_BT$. Insertion and removal of
monomers, with a given orientation, are attempted with equal
probability. For this purpose, an axis orientation (with
probability $1/3$ for the triangular lattice) and a lattice site
are chosen at random. If the selected lattice site is empty,
an attempt to place a monomer (with the orientation previously
chosen) on the site is made. If, instead, the site is occupied,
then the algorithm checks the orientational state of the adsorbed
monomer and if this coincides with the previously chosen
orientation, an attempt to desorb the particle is performed;
otherwise, the trial ends. A MCS is achieved when $M$ sites have
been tested to change its occupancy state. The equilibrium state
can be well reproduced after discarding the first $6 \times 10^8$
MCS. Then, averages are taken over $6 \times 10^8$ successive
configurations.

\subsubsection{Order parameters, Binder cumulant and FSS analysis}

In order to follow the formation of the nematic phase from the
isotropic phase, we use the order parameter defined in Ref.
\cite{Lopez1} for the square lattice,
\begin{equation}
Q =  \frac{\left | N_v -N_h \right |}{N_v +N_h},
 \label{fi}
\end{equation}
where $N_h(N_v)$ is the number of monomers aligned along the
horizontal (vertical) direction, and $N$ is the number of
total monomers on the lattice ($N=N_h + N_v$).

For the triangular and honeycomb lattices, the order parameter
has been defined in Ref. [\onlinecite{Lopez2}] as:

\begin{equation}
Q =  \frac{\left | \vec{n}_1 + \vec{n}_2 + \vec{n}_3 \right
|}{\left | \vec{n}_1 \right | + \left | \vec{n}_2 \right |+ \left
| \vec{n}_3 \right |},
 \label{fi2}
\end{equation}

\noindent where each vector $\vec{n}_i$ is associated to one of the $3$
possible orientations (or directions) for a chain on the lattice.
In addition: $1)$ the $\vec{n}_i$'s lie in the same plane (or are
co-planar) and point radially outward from a given point $P$ which
is defined as coordinate origin; $2)$ the angle between two
consecutive vectors, $\vec{n}_i$ and $\vec{n}_{i+1}$, is equal to
$2\pi/3$; and $3)$ the magnitude of $\vec{n}_i$ is equal to the
number of rods aligned along the $i$-direction (for a graphical
representation, see Fig. 1(a) in Ref. [\onlinecite{Lopez2}]). Note that
the $\vec{n}_i$'s have the same directions as the $q$ vectors in
Ref. [\onlinecite{WU}].

In our canonical MC simulations, we fixed the density $\theta$, and
monitored the order parameter $\langle Q \rangle$ as function of
temperature $T$. The reduced fourth-order (Binder) cumulant $U_L$
\cite{BINDER}, was calculated as:
\begin{equation}
U_L(T) = 1 -\frac{\langle Q^4\rangle} {3\langle
Q^2\rangle^2}, \label{cum}
\end{equation}
\noindent where the thermal average  $\langle ... \rangle$
 means the usual time average throughout the MC
simulation.

The critical behavior of the SAARs model has been investigated by
means of FSS analysis. The FSS theory implies the following
behavior for $\langle Q \rangle$ and $U_L$ at criticality:
\begin{equation}
\langle Q \rangle = L^{-\beta/\nu} \tilde Q(L^{1/\nu} \epsilon),
\label{ds}
\end{equation}
and
\begin{equation}
U_L=\tilde U_L(L^{1/\nu} \epsilon), \label{uls}
\end{equation}
\noindent for $L \rightarrow \infty$, $\epsilon \rightarrow 0$
such that $L^{1/\nu} \epsilon $= finite, where $\epsilon \equiv
T/T_c - 1$ for canonical MC simulations and $\epsilon \equiv
\mu/\mu_c - 1$ for grand canonical MC simulations. Here $\beta$
and $\nu$ are the standard critical exponents of the order
parameter ($\langle Q \rangle \sim (-\epsilon)^{\beta} $ for
$\epsilon\rightarrow 0^-$, $L\rightarrow \infty$), and correlation
length $\xi$ ($\xi \sim |\epsilon|^{-\nu}$ for $\epsilon
\rightarrow 0, L \rightarrow \infty$), respectively. $\tilde Q $
and $\tilde U_L$ are scaling functions for the respective
quantities.

Finally, we calculated the average rod length on the transition
line that, at fixed coverage, increases as the temperature
decreases. At each MCS the average rod length may be
calculated as

\begin{equation}
\overline\ell_{MCS} = \frac{N} {N -\left[N_{bonds} -N_{(L)rods}\right]},
\label{lll}
\end{equation}

\noindent where $N$ is the number of monomers adsorbed on the lattice;
$N_{bonds}$ is the number of bonds between pairs of nearest-neighbors monomers.
 $N_{(L)rods}$ is the number of rods with length $L$; its inclusion prevents counting spurious
bonds introduced by the periodic boundary conditions (i.e., bonds that do not contribute to the rod's length).
Thus the equilibrium average rod length is obtained from

\begin{equation}
\overline\ell = {\langle \overline\ell_{MCS} \rangle},
\label{lllm}
\end{equation}

\subsection{Computational results}

\begin{figure}[t]
\includegraphics[width=8cm,clip=true]{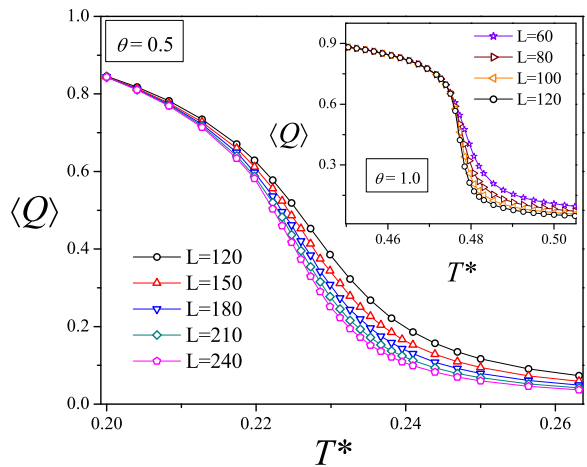}
\caption{Size dependence of the order parameter as a function of
temperature for $\theta= 0.5$ and $\theta = 1$ (inset). }
\label{figure1}
\end{figure}

\subsubsection{Behavior of the Binder cumulant}

The critical behavior of the SAARs model has been investigated by
means of the computational scheme described in the previous
section for the canonical ensemble and FSS analysis
\cite{BINDER,PRIVMAN}. In order to illustrate the behavior of the
Binder cumulant in the critical regime, we show here the results
for the triangular lattice case at intermediate and full coverage.
In addition, these results will be useful when we address the
question of whether or not the universality class can change as a
result of the constant density constraint applied in the canonical
ensemble (Sec. V).

Figure 1 (inset of Fig. 1) shows the behavior of the order parameter versus the reduced temperature $T^*=k_B T/ w$ for several lattice
sizes and $\theta=0.5$ \cite{foot1} ($\theta=1$). As it can be observed, $\langle Q \rangle$ appears as a proper order parameter to elucidate the phase
transition. When the system is disordered ($T^*>T^*_c$, being $T^*_c$
the critical temperature), all orientations are equivalent and
$\langle Q \rangle$ is zero. In the critical regime ($T^*<T^*_c$), the particles
align along one direction and $\langle Q \rangle$ is different from zero.

\begin{figure}[t]
\includegraphics[width=8cm,clip=true]{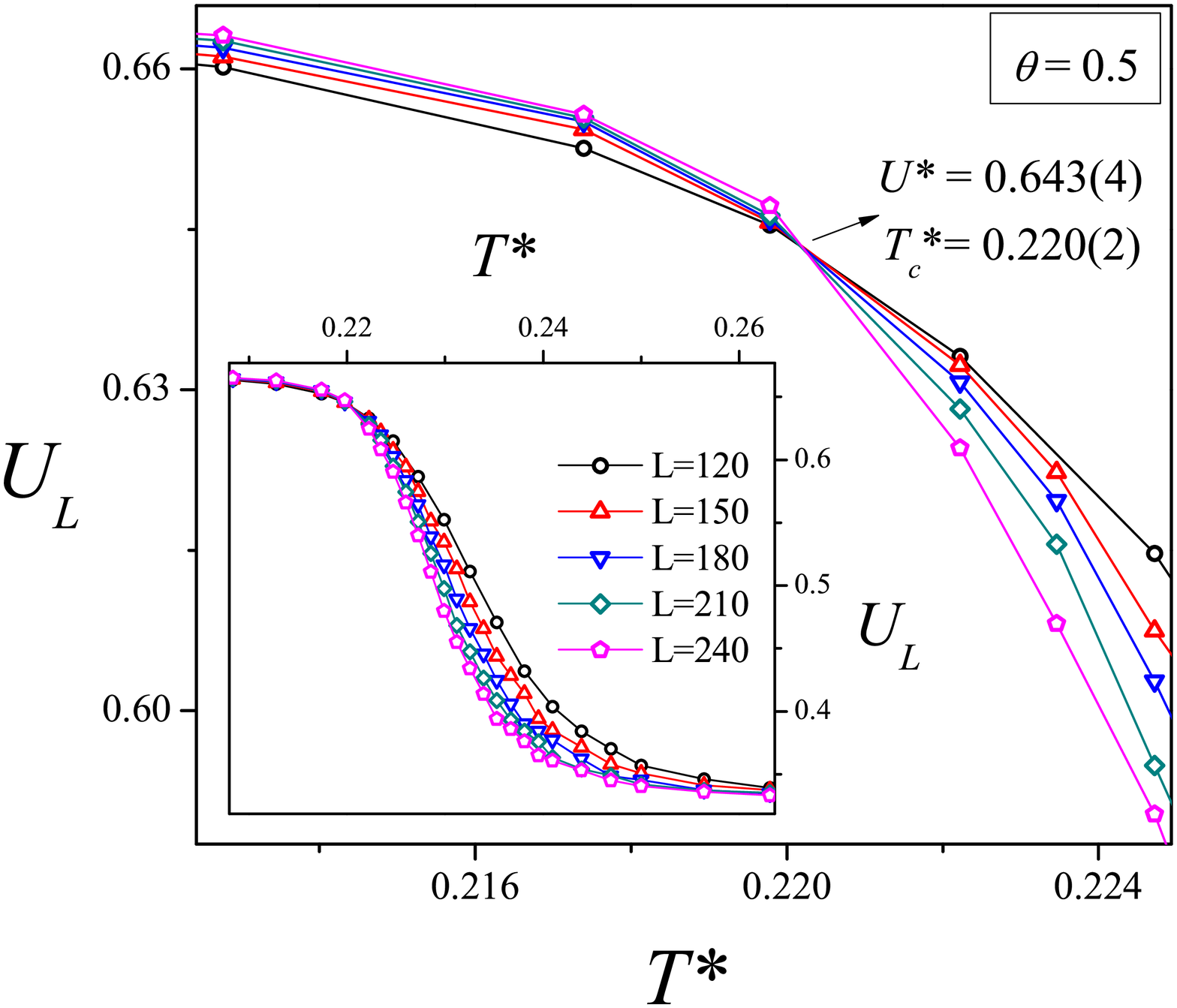}(a)
\includegraphics[width=8cm,clip=true]{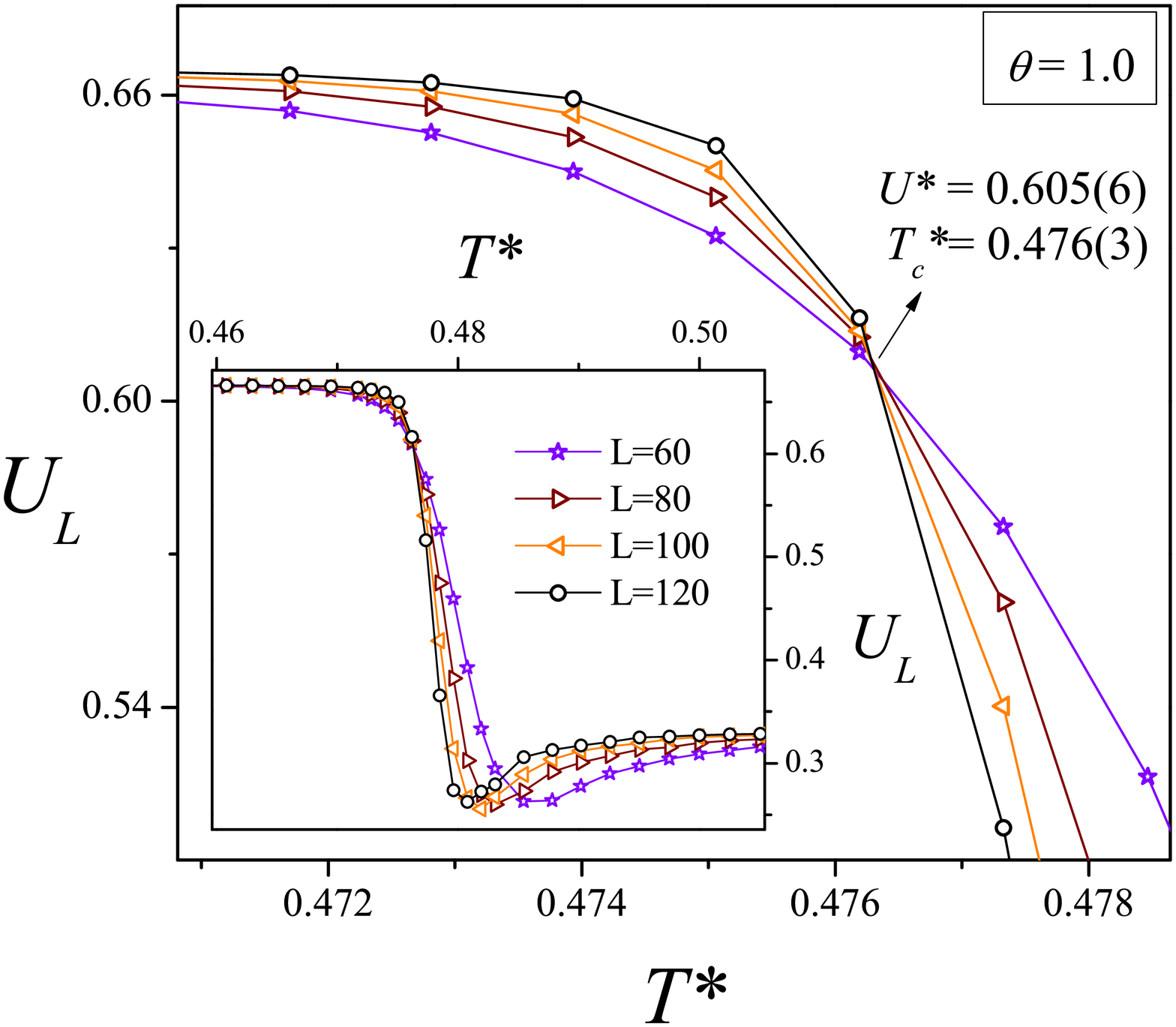}(b)
\caption{Curves of $U_L$ vs $T^*$ for $\theta= 0.5$ (a) and
$\theta  = 1$ (b). From their intersections one obtained $T^*_c$.
In the insets, the data are plotted over a wider range of
temperatures. } \label{figure2}
\end{figure}

Hereafter we discuss the behavior of the critical temperature as a
function of coverage. The standard theory of FSS allows for
various efficient routes to estimate $T^*_c$ from MC data
\cite{BINDER,PRIVMAN}. One of these methods, which will be used in
this case, is from the temperature dependence of $U_L(T^*)$, which
is independent of the system size at $T^*=T^*_c$. In other words,
$T^*_c$ can be found from the intersection of the curve $U_L(T^*)$
for different values of $L$, since $U^* \equiv U_L(T^*_c)=$const.
As an example, Fig. 2 shows the reduced four-order cumulants $U_L$
plotted versus $T^*$ for the cases studied in Fig. 1. The values
obtained for the critical temperature were $T^*_c = 0.220(2)$
(corresponding to $\theta=0.5$) and $T^*_c = 0.476(3)$
(corresponding to $\theta=1$).

The behavior of the reduced fourth-order cumulant as a function of temperature also allows to make a preliminary
identification of the order and universality class of the phase
transition occurring in the system \cite{BINDER}. Thus, the curves in Fig. 2 exhibit the
typical behavior of the cumulants in the presence of a continuous
phase transition. Namely, the order parameter cumulant shows a
smooth drop from $2/3$ to $0$, instead of a characteristic deep
(negative) minimum, as in a first-order phase transition
\cite{BINDER}.

The value of the intersection point $U^*$ shows two
different behaviors, which can be visualized from Fig. 2. On one hand, the value obtained for $U^*$ at
$\theta = 0.5$ ($U^* =0.643(4)$)
is consistent with the $q=1$ Potts universality class \cite{WU}
observed in Ref. [\onlinecite{Lopez2}], where the system was studied
at a fixed temperature ($T^*\approx 0.222$). On the other hand,
at $\theta=1$, the fixed value of the cumulants, $U^* = 0.605(6)$,
is more consistent with previous estimates for the three-state Potts model
(see for instance Ref. [\onlinecite{TOME}], where $U^* \cong 0.613$ \cite{foot2}).
However, even though the value of $U^*$ may be taken as a first indication
of universality, a detailed calculation of critical exponents is
required for an accurate determination of the universality class.
In Sec. V, the distinction between the two universality classes
above is considered based on the determination of the critical exponent
of the correlation length.

\subsubsection{Phase diagrams of SAARs on square, triangular and honeycomb lattices}

\begin{figure}[t]
\includegraphics[width=9.0cm,clip=true]{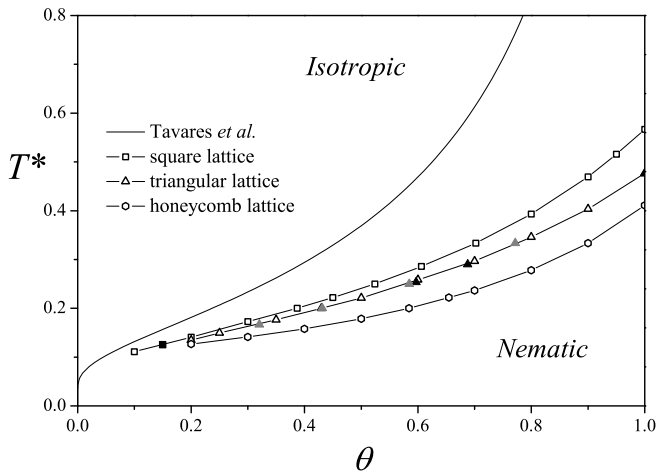}
\caption{Temperature-coverage diagrams for the SARRs model on
different lattices. For comparative purposes, the critical curve
reported by Tavares et al. \cite{Tavares} is shown as a solid
line. The open and solid squares are from Refs.
[\onlinecite{Lopez3}] and [\onlinecite{Almarza1}], respectively.
The open triangles and hexagons represent the results obtained in
this work for the triangular and honeycomb lattices, respectively.
Solid triangles are from Almarza et al. \cite{Almarza2} (in black)
and from Ref. [\onlinecite{Lopez5}] (in grey).} \label{figure3}
\end{figure}

The isotropic-nematic phase transition in a model of self-assembled
rigid rods with restricted orientations was considered for the first
time by Tavares et al. \cite{Tavares}. The temperature-coverage phase
diagram in Ref. [\onlinecite{Tavares}], obtained using an approach
in the spirit of the Zwanzig model \cite{Zwanzig}, is qualitative only,
and the theory overestimates the critical temperature in all ranges
of coverage (especially at high coverages) \cite{foot3}. However, for small
values of $\theta$, small differences appear between simulation and
theoretical results (Fig. 3). As seen in Fig. 3, the critical lines
separate regions of isotropic and nematic stability, and show that
the nematic phase is stable at low temperatures and high densities.
In addition, the phase diagrams show a marked increase of the density difference
from dilute isotropic to dense nematic phases upon increasing
the attraction between monomer units (i.e., decreasing the temperature).

The critical line for SARRs on the square lattice, was obtained by
means of numerical simulations in Refs. [\onlinecite{Lopez3}] and
[\onlinecite{Almarza1}]. Fig. 3 shows the critical line reported
in \cite{Lopez3} and only one point (the lowest coverage obtained)
of the critical line reported in \cite{Almarza1}. As explained in
Ref. [\onlinecite{Almarza1}], the line of critical points
continues beyond the lowest density showed in Fig. 3, however, the
rapid increase of the average length of the rods at low densities
and temperatures prevents an efficient simulation of the system.

For the case of the triangular lattice, some critical points were obtained,
at high coverages in Ref. [\onlinecite{Almarza2}] and at intermediate
coverages in Ref. [\onlinecite{Lopez5}] (see Fig. 3). In the latter case,
the points were obtained from the singularities in the adsorption isotherms.
To corroborate these previous results and complete the phase diagram construction,
the procedure used in III.B.1 (to obtain the critical temperature) was repeated
for $\theta$ ranging between $0.2$ and $1$. The same procedure was done for the case of the honeycomb
lattice, this way the complete phase diagram of honeycomb lattice is reported
here for the first time. All results are collected in Fig. 3. Together,
the phase diagrams show that the critical properties coincide in the
very low-temperature (coverage) regime.

\subsubsection{Critical average rod lengths}

Using a generalization of the theory of associating fluids,
Tavares et al. \cite{Tavares} obtained an analytic expression for
the equilibrium average rod length in the SARRs model, with
orientation $\alpha$ along the $x_1, x_2$ directions in two
dimensions (2D). In Fig. 4 (inset), we present a comparison
between theoretical \cite{Tavares} and numerical results for the
equilibrium average rod length on the transition line. In the
numerical case, the points were obtained by using the Eqs.
(\ref{lll}) and (\ref{lllm}) and square lattices. By comparing
these curves it can be seen a qualitative agreement up to $\theta
\approx 0.8$. However, at higher coverages, MC simulations show a
gradual increase of the critical average rod length, to a value of
$\overline\ell \approx 7$ (at $\theta = 1$). This value is
interesting because it coincides with the minimum value of $k$
($k_{min} = 7$), which allows the formation of a nematic phase in
long straight rigid rods of length $k$ ($k$-mers and monodisperse case), on a square or
triangular lattice \cite{GHOSH,Matoz,Matoz1}.

The average rod length on the critical line was also obtained for
triangular and honeycomb lattices, see Fig. 4. Two observations can be
made from Fig. 4: (i) At intermediate coverage ($\theta
\approx 0.5$, dashed line), the critical average rod lengths for
the square and triangular lattices are similar and near to $7$
(the minimum value of $k$ which allows the formation of a nematic
phase in monodisperse rigid rods, on a square or triangular
lattice \cite{GHOSH,Matoz,Matoz1}). On the other hand, the
critical average rod length (at $\theta \approx 0.5$) for the
honeycomb lattice, is near to $11$, which coincides with the
minimum value of $k$ for the existence of a nematic phase in the case of
monodisperse rigid rods on the honeycomb lattice \cite{Matoz2}.
(ii) Although the trend is more clear for the square and
triangular lattices, all lines tend to converge as the coverage
decreases towards zero; revealing a one-dimensional behavior in
the very low-temperature (coverage) regime. As was shown in Ref.
[\onlinecite{Almarza2}], at zero density limit, where the
average rod length diverges, an equilibrium polymerization
transition occurs.

\begin{figure}[t]
\includegraphics[width=9.0cm,clip=true]{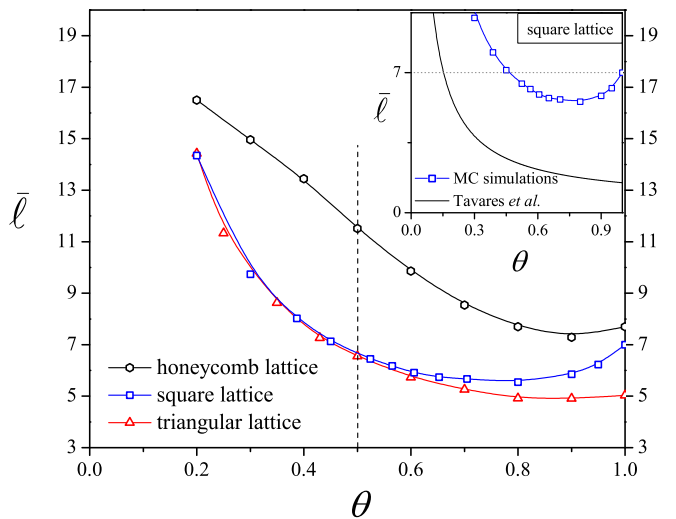}
\caption{ Critical average rod lengths. See description in the text.} \label{figure4}
\end{figure}

\section{ANALYTICAL APPROXIMATIONS}

In this section we calculate  the phase diagram  within the Bethe-Peierls (BP) or quasichemical approximation. To do that we use the Cluster Variation Method (CVM) [\onlinecite{Tanaka}]. In the CVM the BP approximation is obtained minimizing a variational free energy expressed in terms of reduced probability densities, namely

\begin{eqnarray}
     F &=& {\rm Tr} \rho H + k_BT\left\{(1-q_c) \sum_i {\rm Tr}_i  \rho_i^{(1)} \log \rho_i^{(1)} + \right.\nonumber\\
      & & \left. + \sum_{<i,j>}  {\rm Tr}_{i,j}  \rho_{i,j}^{(2)} \log \rho_{i,j}^{(2)} \right\}, \label{BPF}
\end{eqnarray}

\noindent where $q_c$ is the coordination number of the lattice, $\rho_i^{(1)}$ and $\rho_{i,j}^{(2)}$ are one and two sites reduced densities respectively and it is assumed that $\rho = \prod_{<i,j>} \rho_{i,j}^{(2)}$. $\rho_i^{(1)}$ and $\rho_{i,j}^{(2)}$ can be expressed in terms of local one and two site averages, which are used as variational parameters and are related through the reducibility conditions:

\begin{eqnarray}
  {\rm Tr}_i \rho_{i,j}^{(2)} &=& \rho_j^{(1)}, \nonumber\\
  {\rm Tr}_j \rho_{i,j}^{(2)} &=& \rho_i^{(1)}, \label{reduce}
\end{eqnarray}

\noindent and ${\rm Tr}_i \rho_i^{(1)}=1$. For details on the method see, e.g. Ref. [\onlinecite{Tanaka}].
In the next subsections we apply the formalism for the models (\ref{HS1}) and (\ref{Hpotts}).

\subsection{BP approximation for the square lattice case}

Within the representation of the model given by Hamiltonian (\ref{HS1}) the orientational order parameter $\langle Q \rangle$ is basically given by  the average ''magnetization''

\begin{equation}
    m = \frac{1}{M} \left\langle \sum_i S_i \right\rangle,
\end{equation}

\noindent while the coverage is given by

\begin{equation}\label{theta1}
    \theta= \frac{1}{M} \sum_i \left< S_i^2 \right>.
\end{equation}

Then, the one site probability densities can be expressed as
(see Supplementary Material):

\begin{equation}\label{rho1i}
    \rho_i^{(1)}(S_i) =  (1-\theta) + \frac{1}{2}\, m S_i + (\frac{3}{2}\, \theta-1) S_i^2,
\end{equation}

\noindent where we have assumed translational invariance. Defining the two-site correlations

\begin{eqnarray}
  x_{ij} &\equiv & \left<S_i S_j\right>= {\rm Tr}_{i,j} S_i S_j \rho_{i,j}^{(2)}, \\
  y_{ij} & \equiv &\left<S_i^2 S_j^2\right>= {\rm Tr}_{i,j} S_i^2 S_j^2 \rho_{i,j}^{(2)}, \\
  z_{ij} & \equiv &  \left<S_i S_j^2\right>= {\rm Tr}_{i,j} S_i S_j^2 \rho_{i,j}^{(2)},\\
  t_{ij} & \equiv &  \left<S_j S_i^2\right>= {\rm Tr}_{i,j} S_j S_i^2 \rho_{i,j}^{(2)},
\end{eqnarray}

\noindent and imposing the reducibility conditions (\ref{reduce}), the two site density functions $\rho_{i,j}^{(2)}$ can be obtained in terms of the variational parameters $m$, $\theta$, $x_{ij}$,  $y_{ij}$, $z_{ij}$ and  $t_{ij}$ (see Supplementary Material). Replacing into Eq.(\ref{BPF}) we obtain an expression for the variational free energy, whose derivatives can be handled by means of symbolic manipulation programs. Although solving of the corresponding saddle point equations is cumbersome (even numerically), they greatly simplify in the high temperature case, i.e., when we consider the disordered solution $m=t_{ij}=z_{ij}=0$ which is {\it isotropic}: $x_{ij}=x$ and $y_{ij}=y$. In that case we obtain (see Supplementary Material):

\begin{equation}\label{BPm0eq1}
\log (1+y-2\theta)= 2\log(\theta-y)-\log(y-x),
\end{equation}

\begin{equation}\label{BPm0eq2}
    x= y\,  {\rm tanh}\left(\frac{\beta w}{4} \right),
\end{equation}

\begin{eqnarray}
\beta \mu &=& -\log(2)-2 \log(1+y-2 \theta )+\nonumber\\
  & & +3 \log\left(\frac{1-\theta}{\theta}\right)+ 2\,\log(y-x),\label{BPm0eq3}
\end{eqnarray}

\noindent Working out Eqs.(\ref{BPm0eq1})-(\ref{BPm0eq3}) shows that there is only one physically meaningful solution. The equilibrium coverage $\theta^*$  is then given by the solution of the implicit equation

\begin{equation}\label{eqhm0}
    e^{-\beta \mu}=F_{-}(\theta^*,a),
\end{equation}

\noindent where

\begin{equation}
    F_{-}(\theta,a) = 2 \left[ \frac{1+y_{-}(\theta,a)-2\theta}{y_{-}(\theta,a)\,(1-a)} \right]^2 \, \left( \frac{\theta}{1-\theta} \right)^3,
\end{equation}

\noindent with $a\equiv  {\rm tanh}(\beta w/4)$ and

\begin{equation}\label{BPm0eq4}
    y_{-}(\theta,a)=  \frac{1}{2a}\left[1-a+2a\theta -  \sqrt{(1-a+2a\theta)^2-4a\theta^2}\right],
\end{equation}

\noindent while the equilibrium values of the correlations are given by $y^*=y_{-}(\theta^*,a)$ and $x^*= a\, y^*$.

To compute the high temperature nematic susceptibility we add to the Hamiltonian (\ref{HS1}) a small external field $B$ conjugated
to $m$. Then, at temperatures above the critical one we can still assume isotropy in the solution (namely, the solution of the saddle point equations that converge to the previous one in the limit $B\to 0$), so that $t_{ij}=z_{ij}=z$ and $m \ll 1$, $z\ll 1$. This leads to saddle point equations which expanded to the lowest order in $B'\equiv \beta B$ give (see Supplementary Material)

\begin{equation}
    3\, \frac{m}{\theta^*} = 4\, \frac{m-z}{\theta^*-y^*} -B' + {\cal O}(m^2,z^2,mz),
\end{equation}

\begin{equation}
    \frac{z}{x^*+y^*} = \frac{m-z}{\theta^*-y^*}  + {\cal O}(m^2,z^2,mz).
\end{equation}

Then, in the linear response regime $m = \chi B'$ and $z = \omega B'$. In the limit $B'\to 0$, $\chi$ is proportional to the nematic susceptibility. From the above equations we obtain

\begin{eqnarray}
  \chi &=& \frac{\theta^*(\theta^*+x^*)}{3x^*-\theta^*}, \\
  \omega &=& \frac{\theta^*(x^*+y^*)}{3x^*-\theta^*}.
\end{eqnarray}

The disordered solution becomes unstable whenever $3x^*=\theta^*$. Replacing this condition into Eqs.(\ref{BPm0eq1})-(\ref{BPm0eq3}) we obtain the critical line:

\begin{equation}\label{BPcrticalh}
    e^{-\beta\mu_c}= \frac{27}{4} \;\frac{3a-1}{1-a}
\end{equation}

\noindent and

\begin{equation}\label{BPcrticaltheta}
    \theta_c= 3 \frac{1-a}{1+3a},
\end{equation}

\noindent or equivalently:

\begin{equation}\label{BPcrticalT}
    T^*_c= \frac{1}{4\, {\rm arctanh}\left(\frac{1}{3}\,\frac{3-\theta}{1+\theta}\right)}.
\end{equation}

\begin{figure}
\begin{center}
\includegraphics[width=9.0cm,clip=true]{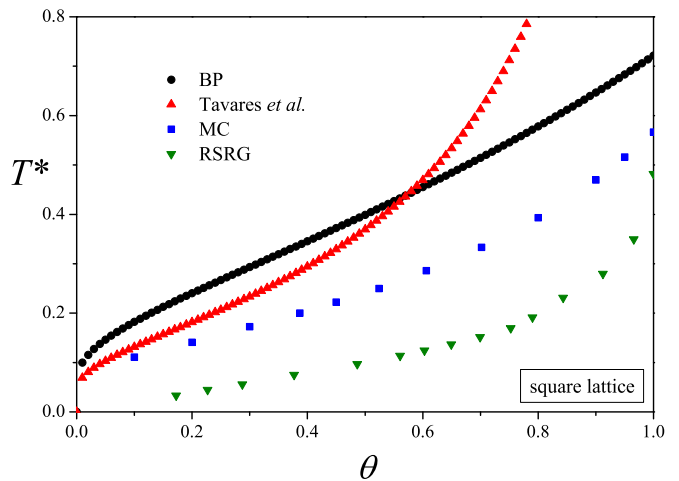}
\caption{\label{fig5} Comparison between the square lattice phase
diagram obtained within the Bethe-Peierls (BP) approximation and
those obtained by other methods: Real Space Renormalization Group
(RSRG) from Ref. [\onlinecite{Lopez3}], Tavares et al. approximation
from Ref. [\onlinecite{Tavares}] and MC simulations.}
\end{center}
\end{figure}

In Fig. \ref{fig5} we compare the Bethe-Peierls critical line with
those obtained by other techniques. From Eq.(\ref{BPcrticalT}) we
obtain the following asymptotic behavior when $\theta \ll 1$:

\begin{equation}\label{BPcrticalTsmalltheta}
    T^*_c\sim -\frac{1}{2\, \ln \left(\theta\right)},
\end{equation}

\noindent which agrees qualitatively with the asymptotic behavior of Tavares et al. calculation $T^*_c \sim -1/3 \ln (\theta)$

\subsection{BP approximation for the triangular lattice case}

The magnetization (orientational order parameter) in this case is given by

\begin{eqnarray}\label{mq}
    m &=& \frac{1}{M} \sum_{i=1}^M \left< \left\{\frac{1}{q-1} \left[q\, \delta(\sigma_i,1)+\delta(\sigma_i,0)-1\right] \right\}\right>\nonumber\\
     &=& \frac{1}{2} \left[3\, \left<\delta(\sigma_i,1)\right>-\theta\right], \label{mq}
\end{eqnarray}

\noindent where the broken symmetry direction $\sigma=1$ is taken arbitrarily among the different $q$ oriented states ($q=3$ in our case). This is a generalization of  usual definition for the $q$-state Potts model. In a disordered state we have $\left<\delta(\sigma_i,1)\right>=\theta/q$, so $m=0$, while in an ordered state along the $\sigma=1$ direction we will have $\left<\delta(\sigma_i,1)\right>=\theta$ so $m=\theta$. A conjugated external field to the order parameter (\ref{mq}) can be considered by adding to the Hamiltonian (\ref{Hpotts}) a term of the form

\begin{equation}
    - \frac{B}{2} \sum_i \left[3\,\delta(\sigma_i,1)+\delta(\sigma_i,0)-1 \right].
\end{equation}

The coverage is given by

\begin{equation}\label{thetaq}
    \theta = \frac{1}{M} \sum_{i=1}^M \left[1- \left<\delta(\sigma_i,0)\right> \right] = 1- \left<\delta(\sigma_i,0)\right>.
\end{equation}

As in the square lattice case, we will consider hereafter only isotropic solutions (valid above the transition temperature in the $B \ll 1$ limit). We then define the correlations

\begin{eqnarray}
  x &\equiv &  \left< \delta(\sigma_i,0)\, \delta(\sigma_j,0)\right>, \\
  y &\equiv & \left< \delta(\sigma_i,1)\, \delta(\sigma_j,1)\right>,\label{BPy1}\\
  z &\equiv& \left< \delta(\sigma_i,2)\, \delta(\sigma_j,2)\right> =\left< \delta(\sigma_i,3)\, \delta(\sigma_j,3)\right>,\label{BPP02}\\
  t &\equiv& \left< \delta(\sigma_i,0)\, \delta(\sigma_j,1)\right>.
\end{eqnarray}

\noindent The one and two sites reduced densities for the  spin variables $\sigma_i=0,1,2,3$ can be expressed as

\begin{equation}
    \rho_i^{(1)}(\sigma_i) = \sum_{\sigma=0}^q  P_\sigma \delta(\sigma_i,\sigma)
\end{equation}

\begin{equation}
    \rho_{i,j}^{(2)}(\sigma_i,\sigma_j)= \sum_{\sigma,\sigma'} P_{\sigma,\sigma'} \delta(\sigma_i,\sigma)\, \delta(\sigma_j,\sigma').
\end{equation}

Applying the normalization and reducibility conditions the coefficients $P_\sigma$ and $P_{\sigma,\sigma'}$ can be expressed in terms of the parameters $m,\theta,x,y,z$ and $t$ (see Supplementary Material for details on the calculation).

First of all we checked the mean-field approximation for this model. Inside the CVM formalism, the mean field free energy can be obtained by assuming that the probability density function is given by $\rho = \prod_i \rho_{i}^{(1)}$ and keeping up to the first order term in the cummulant expansion of the entropy \cite{Tanaka}. With a simple analysis (not shown) we found that the mean field approximation predicts a first order transition for any value of $\mu$, in complete disagreement with the numerical simulations.

Replacing the reduced densities into Eq.(\ref{BPF}) we obtain the BP free energy in terms of the  variational parameters $(m,\theta,x,y,z,t)$ and the corresponding saddle point equations (see Supplementary Material).

At zero field and high enough temperature we have a disordered phase, where all ordered states ($\sigma=1,2,3$) become equally probable and therefore $m=0$ ($\left<\delta(\sigma_i,1)\right>=\theta/q$). Also from the definitions (\ref{BPy1})-(\ref{BPP02}) we have that $y=z$. With some algebra (see Supplementary Material)  the saddle point equations for the disordered solution reduce to

\begin{equation}\label{BPm0SD1}
    \frac{(1-\theta-x)^6 (1-\theta)^5}{x^6 \theta^5}= 3 e^{\beta \mu},
\end{equation}

\begin{equation}\label{BPm0SD2}
    \frac{9zx}{(1-\theta-x)^2}= e^{\beta w/3},
\end{equation}

\begin{equation}\label{BPm0SD3}
   (1-\theta-x)^2 = \frac{3}{2} x\, (2\theta+x-1-3z),
\end{equation}

\begin{equation}\label{BPm0SD4}
   t= (1-\theta-x)/3.
\end{equation}

The physically meaningful solutions of Eqs.(\ref{BPm0SD1})-(\ref{BPm0SD4}) can be obtained in terms of the implicit equation

\begin{equation}
 \theta^* = G_{-}(\theta^*)  \label{G-},
\end{equation}

\noindent where

\begin{eqnarray}
    G_{-}(\theta)&=& \frac{1}{a} \left\{a+ x(\theta) (3-a) -\right. \nonumber\\
     & & \left. - \sqrt{3[ax(\theta)[1-x(\theta)]+3x^2(\theta)]}\right\},\label{Gmn}
\end{eqnarray}

\noindent with $a \equiv 2 + e^{\beta w/3}$ and

\begin{equation}
    x(\theta)= \frac{(1-\theta)^{11/6}}{3^{1/6}\, e^{\beta\mu/6}\, \theta^{5/6}+ (1-\theta)^{5/6}}.\label{BPm0SD4}
\end{equation}

The equilibrium values for the remaining parameters are given by $x^*=x(\theta^*)$, $t^*=(1-\theta^*-x^*)/3$ and

\begin{equation}
    z^*= \frac{e^{\beta w/3}}{9x^*} (1-\theta^*-x^*)^2.
\end{equation}

 Equation (\ref{G-}) has always at least two solutions for any value of $\beta$ and $\mu$: $\theta^*=1$ ($x^*=0$) and $\theta^*=0$ ($x^*=1$). The first one is the meaningful solution in the limit $\mu\to\infty$. For large but finite values of $\mu$ a third solution with $\theta^* <1$ and $1-\theta^* \ll 1$ emerges, which decreases with $\mu$. In the $\mu\ll 1$ we obtain the asymptotic behavior (see Supplementary Material)

 \begin{equation}\label{BPtheta-muinf1}
    1-\theta^* \sim \frac{e^{-\beta\mu}\, 3^5}{(2+e^{\beta w/3})^6},
\end{equation}

\begin{equation}\label{BPtheta-muinf2}
   x^* \sim \frac{e^{-2\beta\mu}\, 3^9}{(2+e^{\beta w/3})^{11}},
\end{equation}

\begin{equation}\label{BPtheta-muinf3}
   z^* \sim \frac{1}{3}\,\frac{e^{\beta w/3}}{2+e^{\beta w/3}}.
\end{equation}

Notice that in the $T\to\infty$ ($\beta\to 0$), the correlation $z^*\to 1/9$, as expected.

At non zero magnetic field $B \ll 1$ we proceeded as in the square lattice case, by expanding the saddle point equations and keeping the lowest order in $B$. This leads to the following expression for the nematic susceptibility (see Supplementary Material):

\begin{equation}\label{BPchi}
    \chi=\frac{1}{2}\; \frac{\theta^* (9z^*-x^*+1)}{12\theta^* -5(9z^*-x^*+1) },
\end{equation}

\noindent which diverges when

\begin{equation}\label{BPtriang-Tc1}
12\theta^* -5(9z^*-x^*+1) =0.
\end{equation}

\begin{figure}
\begin{center}
\includegraphics[width=9.0cm,clip=true]{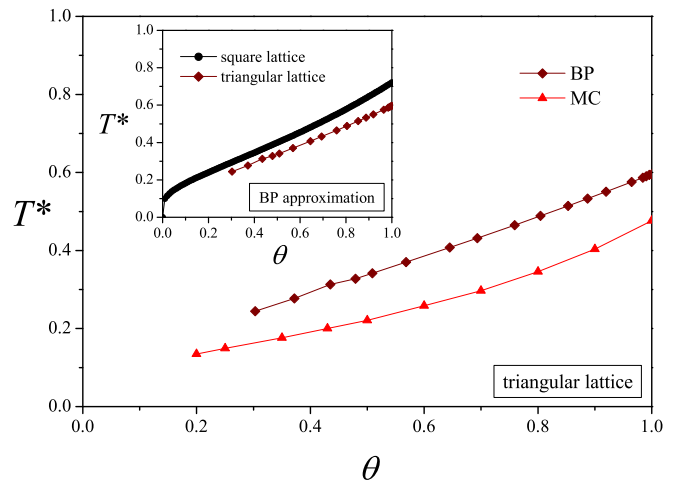}
\caption{\label{fig6} Comparison between the phase diagram
obtained within the Bethe-Peierls (BP) approximation  and MC
simulations for the triangular lattice. Inset: comparison between
the BP critical lines for the square and triangular lattices.}
\end{center}
\end{figure}

For a given value of the chemical potential $\mu/w$,
Eqs.(\ref{Gmn}) and (\ref{BPtriang-Tc1}) must be solved together
in order to obtain the critical line $T^*_c $ {\it
vs.} coverage $\theta$. In Fig.\ref{fig6} we compare a the
critical line obtained by numerically solving Eqs.(\ref{Gmn}) and
(\ref{BPtriang-Tc1}) with the MC results. In particular, in the
limit $\mu\to\infty$, when $\theta\to 1$ and $x\to 0$, we obtain
from Eqs.(\ref{BPtheta-muinf1})-(\ref{BPtheta-muinf3}) that

\begin{equation}\label{BPchi}
    \chi=\frac{1}{4}\; \frac{3+4\, e^{\beta w/3}}{7-4\, e^{\beta w/3} }.
\end{equation}

 We see that $\chi >0$ for all temperatures $T>T_c$, with

\begin{equation}\label{BPtriang-Tc1}
    T^*_c= \frac{1}{3 \log(7/4)} \approx 0.595647.
\end{equation}

\noindent and diverges at that temperature, in a clear signature of a second order phase transition. Comparing with the square lattice result from the previous subsection, we see that the critical temperature at $\theta=1$ decreases in the triangular lattice in a factor $\approx 0.826$, which compares well with the MC reduction factor $\approx 0.841$.

For finite, but relatively large  values of $\mu/w$,
Eqs.(\ref{Gmn}) and (\ref{BPtriang-Tc1}) present only one
nontrivial solution, that converges to the value given by
Eq.(\ref{BPtriang-Tc1}) in the $\mu\to\infty$ limit. However, for
$\mu<\mu_0$, with $\mu_0 \approx -0.65$ (which corresponds to
$\theta \approx 0.73$), two new non trivial solutions emerge, one
with $\theta \ll 1$ and the other with $1-\theta \ll 1$. As $\mu$
further decreases, the low coverage solution approaches the
critical one (i.e., that shown in Fig. \ref{fig6}). Finally, both
solutions collapse at $\mu_c \approx -0.83$ (corresponding to
$\theta\approx 0.24$) and disappear for $\mu < \mu_c$. Such
behavior could be indicative of the presence of a first order
transition at low values of $\mu$, so that the observed secondary
instabilities in the susceptibility would correspond to spinodal
lines. In that case, a tricritical point somewhere along the
calculated transition line should be expected. Indeed, a similar
behavior has been observed within the mean field approximation for
the square lattice \cite{Lopez3}, which disappears in the improved
BP approximation as we have shown in the previous subsection.
However, to check whether there is a change in the transition
order for the triangular lattice case or the observed behavior is
just spurious, requires a complete minimization analysis of the BP
free energy in a multidimensional space (taking anisotropic
ordered solutions into account) which is beyond the scope of the
present work. Whatever the case, the calculation presented in
Fig. \ref{fig6} should be regarded as a high coverage
approximation.

\section{REVISITING THE UNIVERSALITY CLASS}

The purpose of this final section is to revisit a number of the
issues that have emerged during the course of the discussion about
the universality class of the SARRs model, Refs.
[\onlinecite{Lopez1,Almarza1,Almarza2,Lopez4,Almarza3}]. For the
square lattice case \cite{foot4}, at intermediate coverages, it
was shown that the system under study represents an interesting
case where the use of different statistical ensembles (canonical
or grand canonical) leads to different and well-established
universality classes ($q=1$ Potts type or $q=2$ Potts type,
respectively) \cite{Lopez4}. In Ref. [\onlinecite{Almarza3}],
Almarza et al. concluded that the dependence of the universality
class of the SARRs model on the statistical ensemble, is very
likely the result of inadequate use of normal scaling to
investigate the critical properties of the constrained (constant
density) system. However, to date, no completely satisfactory
explanation has been given on the consistency of the FSS behavior,
in the canonical ensemble, with the critical exponents of the
ordinary percolation (i.e. 2D Potts $q=1$ universality class).

As in III.B.1, we will address here only the triangular lattice
case. We expect the same universality class for chains on
honeycomb lattices (with three allowed orientations), given that
the excluded volume term exhibits the same symmetry. The critical
behavior of the SAARs model on a triangular lattice was recently
reinvestigated by Almarza et al. \cite{Almarza2}. The authors
found that the isotropic-nematic phase transition occurring in the
system is in the 2D Potts $q=3$ universality class. This
conclusion contrasts with that of a previous study in the
canonical ensemble \cite{Lopez2} which indicates that the
transition in triangular (and honeycomb) lattices, at intermediate
density, belongs to the q = 1 Potts universality class. In Ref.
[\onlinecite{Almarza2}], Almarza et al. attributed the discrepancy
to the use of the density as the scaling variable in Ref.
[\onlinecite{Lopez2}]. In addition, Almarza et al. have cited a
paper \cite{Fisher} in which Fisher showed that fixing the density
in some models corresponds to introducing a constraint that
renormalizes the critical exponents. More precisely, Almarza et
al. have noted that, for the Potts $q=3$ universality class, the
renormalized correlation length exponent $\nu'$ is
$\nu'=\nu/(1-\alpha)=5/4$, which is close to the value of $\nu$
for the $q=1$ universality class, $\nu_{q=1}=4/3$, reported in
Ref. [\onlinecite{Lopez2}].

\begin{figure}[t]
\includegraphics[width=8.0cm,clip=true]{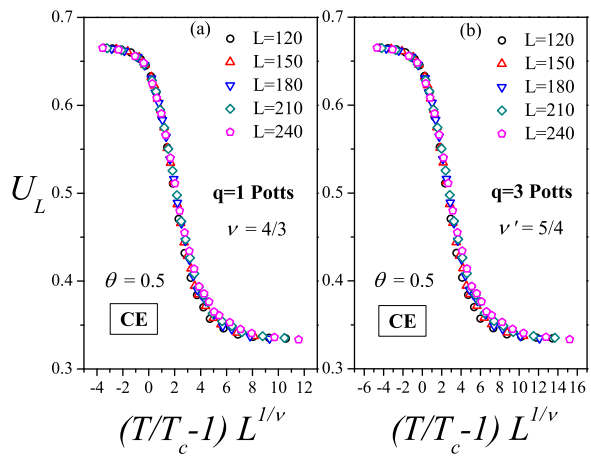}
\caption{ Data collapsing of the Binder cumulant, $U_L$ vs $\epsilon
L^{1/\nu}$, with the correlation length exponent of the ordinary percolation (a), and with
the  renormalized exponent (b). CE means canonical ensemble.} \label{figure5}
\end{figure}

\begin{figure}[t]
\includegraphics[width=8.0cm,clip=true]{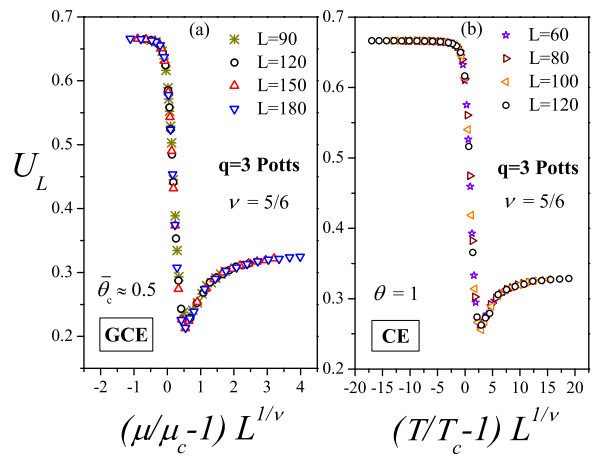}
\caption{ Data collapsing of the Binder cumulant, $U_L$ vs $\epsilon
L^{1/\nu}$, with the correlation length exponent of the 2D Potts $q=3$ universality class,
for: grand canonical ensemble (GCE) simulations at intermediate coverage (a),
and canonical ensemble (CE) simulations at full-coverage (b).} \label{figure6}
\end{figure}

In order to test the argument given by Almarza et al., a series of
MC simulations have been conducted. As in Refs.
[\onlinecite{Almarza1, Lopez4}], the distinction between the two
universality classes is based on the determination of the value of
$\nu$, which is clearly different for the two universality classes
under discussion. Then, the scaling behavior can be tested by
plotting $U_L$ vs $\epsilon L^{1/\nu}$ and looking for data
collapse. As shown in Fig. 7(a), the collapse of the curves
corresponding to the Fig. 2(a), where the control parameter is the
temperature, provides convincing evidence that the scaling
obtained using $\nu_{q=1}=4/3$ is not due to the use of the
density as the control parameter, as claimed by Almarza et al.
However, as would be expected due to the proximity of the values
considered here ($\nu_{q=1}$ and $\nu'$), a good data collapse
with the renormalized correlation length exponent $\nu'$ can also
be obtained [Fig. 7(b)]. Hence, unlike what happens in the square
lattice case, Fisher renormalization arguments appear to be
sufficient in the triangular (honeycomb) lattice case.

Moreover, to check the data presented by Almarza et al.
\cite{Almarza2}, MC simulations in the grand canonical ensemble
were carried out using an adsorption-desorption algorithm. It is
important to note that the algorithm used here is different from
that used by Almarza et al. In the grand canonical ensemble, the
critical behavior was studied at the same point of the phase
diagram ($\overline\theta_c \approx 0.5$), fixing the inverse of
the reduced temperature to $1/T^* = 4.5$, and varying the chemical
potential $\mu$. Very good collapse was obtained with $\nu=5/6$ in
the scaling plot of $U_L$ [Fig. 8(a)], thus corroborating the data
of Almarza et al. In addition, only in the full-lattice limit
($\theta = 1$), canonical MC simulations are able to produce
results consistent with the 2D Potts $q=3$ universality class.
[see Fig. 8(b), which shows the collapse of the cumulant curves
corresponding to the Fig. 2(b)].

Finally, we wish to clarify that: (i) We do not hold that the
universality class of the SARRs model depends on the
polydispersity of the rods, as was stated by Almarza et al.
\cite{Almarza3} in reference to our work \cite{Lopez4}. (ii)  We
agree with Almarza et al. that the universality class of the SARRs
model, in the square lattice, is that of the 2D Ising model,
whereas that in the triangular and honeycomb lattices, is the same
as that of the 2D Potts model with $q=3$. (iii) The strong
consistency of the results obtained in the canonical ensemble with
the critical exponents of the ordinary percolation (at
intermediate coverage, in the three lattices considered), warrants
an explanation that has not yet been given.

\section{CONCLUSIONS}

In this paper, the main critical properties of self-assembled
rigid rods on square, triangular and honeycomb lattices have been
addressed. The results were obtained by using Monte Carlo
simulations in the canonical and grand canonical ensembles,
finite-size scaling techniques and theoretical analysis in the
framework of the Bethe-Peierls approximation.

Several conclusions can be drawn from the present work. On the one
hand, the equilibrium average rod length as a function of
concentration was calculated by MC simulations. In the case of
square lattices, computational data were compared with theoretical
results from Tavares et al. \cite{Tavares}. A good qualitative
agreement was observed in the range of coverage from 0 to 0.8.
However, the disagreement turns out to be significantly large for
$\theta > 0.8$. In the case of triangular and honeycomb lattices,
the dependence of the equilibrium average rod length on coverage
was reported here for the first time.

The obtained results for $\overline\ell(\theta)$ reveal two
interesting observations: (i) at intermediate coverage ($\theta
\approx 0.5$), the value of the average rod length coincides with
the minimum value of $k$ ($k_{min}$), which allows the formation
of a nematic phase for a system of monodisperse straight rigid
$k$-mers adsorbed on two-dimensional lattices (square lattice,
$k_{min}=7$ \cite{GHOSH,Matoz,Matoz1}; triangular lattice,
$k_{min}=7$ \cite{Matoz,Matoz1} and honeycomb lattice,
$k_{min}=11$ \cite{Matoz2}); and (ii) at low coverage, the three
curves show the same tendency, independently of the lattice
geometry (given the range of concentrations studied here, this
behavior is more evident for square and triangular lattices). This
finding reinforces the idea that the adsorption process behaves as
a one-dimensional problem in the low-coverage (temperature)
regime: particles adsorb forming chains and an equilibrium
polymerization transition occurs in the system \cite{Lopez5}.

On the other hand, and regarding the phase diagram of the SARRs,
the complete $T$-$\theta$ critical curves corresponding to
triangular and honeycomb lattices have been obtained by using MC
simulation and FSS analysis. In the case of triangular lattices,
the present study allowed us to corroborate previous results
obtained from the behavior of the adsorption isotherms
\cite{Lopez5} and, in the case of honeycomb lattices, the phase
diagram has been reported here for the first time.

The simulation phase diagrams were compared with analytical data
from the BP approximation. BP results confirm the whole scenario
that emerges from the MC simulations, in a clear improvement
respect to the mean-field approximation, namely: a) a continuous
nature of the phase transition for any value of $\theta$ (although
in the triangular lattice case the BP fails in that feature at low
coverage, the general trend suggests to be an spurious effect of
the approximation); b) a consistent reduction in the critical
temperatures for any value $\theta$ when the triangular and square
lattice models are compared; c) an independence on $\theta$ of the
critical curves for different lattices at very low values of
$\theta$ (see inset of Fig.\ref{fig6}) and d) a logarithmic
decrease with $\theta$ of the critical curve when $\theta\to 0$,
in agreement with other analytical approach \cite{Tavares}.

In an earlier study \cite{Lopez3}, in which the critical behavior
of SARRs on the square lattice was addressed, it was shown that in
the full coverage case ($\theta=1.0$), the Hamiltonian of the
SAARs model maps exactly onto the Ising model one ($q=2$ Potts
model) with coupling constant $w_{Ising}=w_{SARRs}/4$. In
contrast, the 2D SARRs model on the triangular lattice cannot be
mapped on $q=3$ Potts model, as can be easily seen from Eq.
(\ref{Hpotts}). It is interesting to note, that through a simple
extension of the present calculations the critical temperature
$T^*_c$ within the BP approximation for the isotropic $q=3$ Potts
model results $3$ times that of the SARRs on the triangular
lattice. The corresponding comparison between the critical
temperatures extracted from MC simulations predicts a factor
$3,3299 \approx 10/3$, in close agreement with the BP result.

Finally, the problem of the universality class of the SAARs model
was revisited. Since the case corresponding to square lattices has
been widely discussed in Refs.
[\onlinecite{Lopez1,Almarza1,Lopez4,Almarza3}], we focused in the
case of triangular lattices (the same universality class is
expected to hold also for honeycomb lattices with three allowed
orientations). Based on the calculation of the correlation length
exponent $\nu$ in the canonical and grand canonical ensembles, and
using the Fisher renormalization scheme, we confirmed previous
results by Almarza et al. \cite{Almarza2}. Namely, the
universality class of the SARRs model for triangular and honeycomb
lattices is that of the 2D Potts model with $q=3$. However, the
strong consistency of the results obtained in the canonical
ensemble with the critical exponents of the ordinary percolation
(at intermediate coverage, in the three lattices considered),
warrants an explanation that has not yet been given.

\acknowledgments This work was supported in part by CONICET
(Argentina) under projects number PIP 112-200801-01332 and
112-200801-01576; Universidad Nacional de San Luis (Argentina)
under project 322000; Universidad Nacional C\'ordoba (Argentina),
the National Agency of Scientific and Technological Promotion
(Argentina) under project  PICT-2010-1466, Secretar\'{\i}a de
Pol\'{\i}ticas Universitarias  under grant  PPCP007/2012
(Argentina) and CAPES  (Brasil) under grant PPCP 007/2011.




\begin{thebibliography}{99}
\bibitem{foot0} If the operation $div$ is not used in (\ref{hc}),
there will be three possible values of energy between each pair of
nearest-neighbor monomers on the triangular (honeycomb) lattice: $w$ if they
are aligned in the direction of the intermolecular vector
$\vec{r}_{ij}$; $w/2$ if one is aligned in the direction of
$\vec{r}_{ij}$ and the other not; and $w/4$ if they are oriented
in one direction but different to $\vec{r}_{ij}$.
\bibitem{foot2} We have calculated the value of $U^*$ for the
standard $q=3$ Potts model in the triangular lattice (with
periodic boundary conditions), via MC simulation. The value
obtained was $U^* = 0.613(3)$, in excellent agreement with those
reported in Ref. [\onlinecite{TOME}].
\bibitem{foot3} In Ref. \cite{Lopez3}, we have addressed the
temperature-coverage phase diagram of self-assembled rigid rods on
square lattices by using real space renormalization group (RSRG)
approach. The RSRG approach reproduces qualitatively the shape of the
critical line, but systematically underestimates the critical
temperature.
\bibitem{foot4} It must be pointed out that the RSRG approach,
developed in Ref. \cite{Lopez3} for the square lattice, predicts
that the whole critical line is in the universality class of the
$d=2$ ferromagnetic Ising model, in concordance with the results
reported in Ref. \cite{Almarza1}.
\end{thebibliography}
\end{document}


\begin{center}
{\Large \bf Supplementary information}

\vspace{0.2cm}

{\large \bf Critical behavior of self-assembled rigid rods on two-dimensional lattices: Bethe-Peierls approximation and Monte Carlo simulations}

\vspace{0.2cm}

L. G. L\'opez, D. H. Linares, A. J. Ramirez-Pastor, D. A. Stariolo and S. A. Cannas

\end{center}

Here we provide the details of the BP approximation presented in the manuscript.

\section{The BP approximation in the square lattice case}
\label{BPsquare}

 Following the method outlined in Ref.[\onlinecite{Tanaka}] (see chapter 6)
the Bethe-Peierls or two-point cluster free energy for the square lattice model can be written as:

\begin{eqnarray}
    F &=& {\rm Tr} \rho H + k_BT\left\{(1-q_c) \sum_i {\rm Tr}_i  \rho_i^{(1)} \log \rho_i^{(1)} + \sum_{<i,j>}  {\rm Tr}_{i,j}  \rho_{i,j}^{(2)} \log \rho_{i,j}^{(2)} \right\}\nonumber\\
     &=& -\frac{w}{4} \sum_{<i,j>} \left[(x_{ij}+y_{ij}+z_{ij}+w_{ij})({\bf e}_y . \vec{r}_{ij})+  (x_{ij}+y_{ij}-z_{ij}-w_{ij})({\bf e}_x.\vec{r}_{ij}) \right] -\mu \sum_i \theta_i + \nonumber\\
     && + k_BT\left\{(1-q) \sum_i {\rm Tr}_i  \rho_i^{(1)} \log \rho_i^{(1)} + \sum_{<i,j>}  {\rm Tr}_{i,j}  \rho_{i,j}^{(2)} \log  \rho_{i,j}^{(2)} \right\}
\label{BPF}
\end{eqnarray}

where:

\begin{eqnarray} \label{correls}
  x_{ij} &\equiv & \left<S_i S_j\right>= {\rm Tr}_{i,j} S_i S_j \rho_{i,j}^{(2)} \\
  y_{ij} & \equiv &\left<S_i^2 S_j^2\right>= {\rm Tr}_{i,j} S_i^2 S_j^2 \rho_{i,j}^{(2)} \\
  z_{ij} & \equiv &  \left<S_i S_j^2\right>= {\rm Tr}_{i,j} S_i S_j^2 \rho_{i,j}^{(2)}\\
  w_{ij} & \equiv &  \left<S_j S_i^2\right>= {\rm Tr}_{i,j} S_j S_i^2 \rho_{i,j}^{(2)}
\end{eqnarray}

We start calculating the one point and two points reduced densities
$\rho_i^{(1)}(S_i)$ and $\rho_{i,j}^{(2)}(S_i,S_j)$.

In full generality, the one point density for the model considered can be written as:

\begin{equation}
    \rho_i^{(1)}(S_i) = A + B S_i + C S_i^2.
\end{equation}

Imposing

\begin{equation}\label{norm1}
    {\rm Tr}_i \rho_i^{(1)}(S_i) = 1
\end{equation}

\begin{equation}\label{mi}
    m_i \equiv \left<S_i \right>= {\rm Tr}_i S_i \rho_i^{(1)}(S_i)
\end{equation}

\begin{equation}\label{thetai}
    \theta_i \equiv \left<S_i^2 \right>= {\rm Tr}_i S_i^2 \rho_i^{(1)}(S_i)
\end{equation}

the values of the coefficients $A,B$ and $C$ are found to be:

\begin{equation}\label{rho1i}
    \rho_i^{(1)}(S_i) =  (1-\theta_i) + \frac{1}{2}\, m_i S_i + (\frac{3}{2}\, \theta_i-1) S_i^2
\end{equation}

Analogously for the two point density, consider the expansion:

\begin{equation}
    \rho_{i,j}^{(2)}(S_i,S_j)= a + b_i S_i + b_j S_j + c_i S_i^2 + c_j S_j^2 + d\, S_iS_j + e\, S_i^2 S_j^2 + f_{ij} S_i S_j^2+ f_{ji}  S_j S_i^2
\end{equation}

\noindent Imposing the reducibility conditions

\begin{eqnarray}
  {\rm Tr}_i \rho_{i,j}^{(2)} &=& \rho_j^{(1)} \label{reduce1}\\
  {\rm Tr}_j \rho_{i,j}^{(2)} &=& \rho_i^{(1)} \label{reduce2}
\end{eqnarray}

\noindent the following set of equations is obtained:

\begin{eqnarray}
  3a+2c_i &=&  (1-\theta_j)\label{aa1} \\
  3a+2c_j &=&  (1-\theta_i)\label{aa2} \\
  3b_i+2f_{ij} &=&  \frac{1}{2}\, m_i\label{aa3}\\
  3b_j+2f_{ji} &=&  \frac{1}{2}\, m_j\label{aa4}\\
  3c_i+2e &=&  \frac{3}{2}\, \theta_i-1\label{aa5} \\
  3c_j+2e &=&  \frac{3}{2}\, \theta_j-1 \label{aa6}
\end{eqnarray}

\noindent Actually, these equations are not all linearly independent, since replacing (\ref{aa5}) into (\ref{aa2}) and (\ref{aa6}) into (\ref{aa1}) we obtain the same equation (Eq.(\ref{aa8})). Hence, we can choose the following set of linearly independent equations

\begin{eqnarray}
  3a-\frac{4}{3}\, e &=& \frac{5}{3} -\theta_i-\theta_j\label{aa7} \\
  3b_i+2f_{ij} &=&  \frac{1}{2}\, m_i\label{aa8}\\
  3b_j+2f_{ji} &=&  \frac{1}{2}\, m_j\label{aa9}\\
  3c_i+2e &=&   \frac{3}{2}\, \theta_i-1\label{aa10} \\
  3c_j+2e &=&  \frac{3}{2}\, \theta_j-1  \label{aa11}
\end{eqnarray}

With the definitions (\ref{correls}) we find $d=x_{ij}/4$. Also,

\begin{eqnarray}
  y_{ij} &=& 4a + 4 (c_i+c_j) +4e\label{aa12} \\
  z_{ij} &=&  4 b_i +4f_{ij}\label{aa13} \\
  w_{ij} &=&  4 b_j +4f_{ji}\label{aa14}
\end{eqnarray}

\noindent From Eqs.(\ref{aa8}), (\ref{aa9}), (\ref{aa13}) and (\ref{aa14}) we obtain

\begin{eqnarray}
  b_i &=& \frac{1}{2}(m_i - z_{ij}) \\
  b_j &=& \frac{1}{2}(m_j - w_{ij})
\end{eqnarray}

\begin{eqnarray}
  f_{ij} &=& \frac{3}{4}\, z_{ij} -\frac{1}{2}\, m_i \\
  f_{ji} &=& \frac{3}{4}\, w_{ij} -\frac{1}{2}\, m_j
\end{eqnarray}

\noindent From Eqs.(\ref{aa7}), (\ref{aa10}), (\ref{aa11}) and (\ref{aa12}) we obtain

\begin{eqnarray}
  c_i &=& -1 -\frac{3}{2}\, y_{ij} + \frac{3}{2}\,\theta_i+\theta_j \\
  c_j &=& -1 -\frac{3}{2}\, y_{ij} + \frac{3}{2}\,\theta_j+\theta_i
\end{eqnarray}

\begin{equation}
    a = y_{ij} + 1-(\theta_i+\theta_j)
\end{equation}

\begin{equation}
    e =\frac{9}{4}\, y_{ij} + 1 -\frac{3}{2}(\theta_i+\theta_j)
\end{equation}

Now, from the symmetries of Hamiltonian   (Eq.(3) of the manuscript) we can assume $m_i=m$, $\theta_i=\theta$, but the correlations $x_{ij}$, $y_{ij}$ and $w_{ij}$ and $z_{ij}$ may depend on orientation along the principal axes of the lattice. Therefore

\begin{equation}\label{rho1i2}
    \rho^{(1)}(S_i) =  (1-\theta) + \frac{1}{2}\, m S_i + \left(\frac{3}{2}\, \theta-1\right) S_i^2
\end{equation}
and
\begin{eqnarray}
    \rho_{i,j}^{(2)}(S_i,S_j)&=& (y_{ij}+1-2\theta ) + \frac{1}{2}(m-z_{ij}) S_i +
\frac{1}{2}(m-w_{ij}) S_j + \left[-1-\frac{3}{2}\, y_{ij} +\frac{5}{2} \theta\right] (S_i^2 +  S_j^2) + \nonumber\\
    & + & \frac{x_{ij}}{4}\, S_iS_j + \left[\frac{9}{4}\,y_{ij}+1-3\theta \right]\, S_i^2 S_j^2 + \left[\frac{3}{4}\,z_{ij}-\frac{1}{2}\, m \right] S_i S_j^2+  \left[\frac{3}{4}\,w_{ij}-\frac{1}{2}\, m \right] S_j S_i^2.
\end{eqnarray}
Writing $x_{ij}=x_{\|}$ and $x_{ij}=x_{\perp}$ for pair correlations along ${\mathbf e}_x$ and
${\mathbf e}_y$ respectively, and the same for $y_{ij}$, $w_{ij}$ and $z_{ij}$, we can write the
variational free energy (\ref{BPF}) as:

\begin{eqnarray}\label{FBP}
    \beta F/N & = & -\frac{K}{4}\left[ (x_{\|}+x_{\perp})+(y_{\|}+y_{\perp})-(z_{\|}-z_{\perp})-(w_{\|}-w_{\perp})\right] -h\theta \nonumber \\
 & & + (1-q)  \sum_{S=0,\pm 1} \rho^{(1)}(S) \log \rho^{(1)}(S) \\
& & +  \sum_{S_1=0,\pm 1}\sum_{S_2=0,\pm 1}   \rho_{1,2}^{(2)}(S_1,S_2) \log  \rho_{1,2}^{(2)}(S_1,S_2)
\left[({\mathbf e}_x\cdot \vec r_{12})+({\mathbf e}_y\cdot \vec r_{12})\right].\nonumber
\end{eqnarray}

\noindent where $K\equiv\beta w$. The last trace has contributions from horizontal and vertical links. Calling
 $\rho_{1,2}^{(2)}(S_1,S_2) = \rho_{1,2}^{(2),para}(S_1,S_2)$ for the horizontal case:
\begin{eqnarray}
    \rho_{1,2}^{(2),para}(S_i,S_j)&=& (y_{\|}+1-2\theta ) + \frac{1}{2}(m-z_{\|}) S_i +
\frac{1}{2}(m-w_{\|}) S_j + \left[-1-\frac{3}{2}\, y_{\|} +\frac{5}{2} \theta\right] (S_i^2 +  S_j^2) + \nonumber\\
    & + & \frac{x_{\|}}{4}\, S_iS_j + \left[\frac{9}{4}\,y_{\|}+1-3\theta \right]\, S_i^2 S_j^2 + \left[\frac{3}{4}\,z_{\|}-\frac{1}{2}\, m \right] S_i S_j^2+  \left[\frac{3}{4}\,w_{\|}-\frac{1}{2}\, m \right] S_j S_i^2,
\end{eqnarray}
and a similar expression for the vertical terms $\rho_{1,2}^{(2),perp}(S_1,S_2)$, we arrive at a
rather long expression for the free energy, which can be conveniently handled by a software
for symbolic manipulation like Mathematica.

\subsection{High temperature disordered solution}

In the high temperature phase $m=w=z=0$ and correlations are isotropic. Deriving the
variational free energy (\ref{BPF}) with respect to $\theta$ we find:

\beq
h=-2 \text{Log}[1+y_{\|}-2 \theta ]-2 \text{Log}[1+y_{\perp}-2 \theta ]+2 \text{Log}\left[-\frac{y_{\|}}{2}+\frac{\theta }{2}\right]+2
\text{Log}\left[-\frac{y_{\perp}}{2}+\frac{\theta }{2}\right]-3 \left(-\text{Log}[1-\theta ]+\text{Log}\left[\frac{\theta }{2}\right]\right)
\eeq

Deriving respect to $x_{\|}$ :
\beqa \label{sp1}
\frac{K}{4} &=& -\frac{1}{2} \text{Log}\left[2-\frac{x_{\|}}{4}+\frac{13 y_{\|}}{4}-5 \theta +2 \left(-1-\frac{3 y_{\|}}{2}+\frac{5
\theta }{2}\right)\right] \\
& + & \frac{1}{2} \text{Log}\left[2+\frac{x_{\|}}{4}+\frac{13 y_{\|}}{4}-5 \theta +2 \left(-1-\frac{3 y_{\|}}{2}+\frac{5
\theta }{2}\right)\right]. \nonumber
\eeqa

Deriving respect to $x_{\perp}$ :
\beqa \label{sp2}
\frac{K}{4} &=& -\frac{1}{2} \text{Log}\left[2-\frac{x_{\perp}}{4}+\frac{13 y_{\perp}}{4}-5 \theta +2 \left(-1-\frac{3 y_{\perp}}{2}+\frac{5
\theta }{2}\right)\right] \\
& +&\frac{1}{2} \text{Log}\left[2+\frac{x_{\perp}}{4}+\frac{13 y_{\perp}}{4}-5 \theta +2 \left(-1-\frac{3 y_{\perp}}{2}+\frac{5
\theta }{2}\right)\right]. \nonumber
\eeqa

Deriving respect to $y_{\|}$ :
\beqa \label{sp3}
\frac{K}{4}& =&\text{Log}[1+y_{\|}-2 \theta ]-2 \text{Log}\left[-\frac{y_{\|}}{2}+\frac{\theta }{2}\right]+\frac{1}{2} \text{Log}\left[2-\frac{x_{\|}}{4}+\frac{13
y_{\|}}{4}-5 \theta +2 \left(-1-\frac{3 y_{\|}}{2}+\frac{5 \theta }{2}\right)\right] \nonumber \\
& & +\frac{1}{2} \text{Log}\left[2+\frac{x_{\|}}{4}+\frac{13
y_{\|}}{4}-5 \theta +2 \left(-1-\frac{3 y_{\|}}{2}+\frac{5 \theta }{2}\right)\right].
\eeqa

Deriving respect to $y_{\perp}$ :
\beqa \label{sp4}
\frac{K}{4}&=& \text{Log}[1+y_{\perp}-2 \theta ]-2 \text{Log}\left[-\frac{y_{\perp}}{2}+\frac{\theta }{2}\right]+\frac{1}{2} \text{Log}\left[2-\frac{x_{\perp}}{4}+\frac{13
y_{\perp}}{4}-5 \theta +2 \left(-1-\frac{3 y_{\perp}}{2}+\frac{5 \theta }{2}\right)\right]
\nonumber \\
 & & +\frac{1}{2} \text{Log}\left[2+\frac{x_{\perp}}{4}+\frac{13
y_{\perp}}{4}-5 \theta +2 \left(-1-\frac{3 y_{\perp}}{2}+\frac{5 \theta }{2}\right)\right].
\eeqa

We see that eqa. (\ref{sp1}) and (\ref{sp2}) are equal and the same happens with
(\ref{sp3}) and (\ref{sp4}). This confirms that the disordered solution is isotropic
with respect to correlations and we can write $x_{\|}=x_{\perp}=x$ and $y_{\|}=y_{\perp}=y$.

The previous set of equations can be conveniently reduced to:

\begin{equation}
h= -\log(2)-4 \log(1+y-2 \theta )+3 \log(1-\theta )-3 \log(\theta )
 + 4 \log(-y+\theta) \label{BPtheta1}
\end{equation}

\begin{eqnarray}
\frac{K}{2} &=& -\log \left[2-\frac{x}{4}+\frac{13 y}{4}-5 \theta +2 \left(-1-\frac{3 y}{2}+\frac{5\theta }{2}\right)\right] \nonumber \\
& & + \log \left[2+\frac{x}{4}+\frac{13 y}{4}-5 \theta +2 \left(-1-\frac{3 y}{2}+\frac{5
\theta }{2}\right)\right] \nonumber \\
        &=& \log (y+x) - \log (y-x)
\end{eqnarray}

\begin{eqnarray}
\frac{K}{2}& =& 2 \log [1+y-2 \theta ]-4 \log \left[-\frac{y}{2}+\frac{\theta }{2}\right]+
\log \left[2-\frac{x}{4}+\frac{13
y}{4}-5 \theta +2 \left(-1-\frac{3 y}{2}+\frac{5 \theta }{2}\right)\right] \nonumber \\
& & + \log \left[2+\frac{x}{4}+\frac{13y}{4}-5 \theta +2 \left(-1-\frac{3 y}{2}+
\frac{5 \theta }{2}\right)\right] \nonumber \\
 &=&  \log (y-x) +\log (y+x)  +2\log (1+y-2\theta) -  4 \log(\theta-y)
\end{eqnarray}

\noindent where $h\equiv\beta\mu$. These equations can be further simplified to give:

\begin{equation}\label{BPm0eq1}
\log (1+y-2\theta)= 2\log(\theta-y)-\log(y-x)
\end{equation}

\begin{equation}\label{BPm0eq2}
    x= y\,  {\rm tanh}\left(\frac{K}{4} \right)
\end{equation}

\begin{equation}
h= -\log(2)-2 \log(1+y-2 \theta )+3 \log\left(\frac{1-\theta}{\theta}\right)+ 2\,\log(y-x)\label{BPm0eq3}
\end{equation}

\noindent Combining Eqs.(\ref{BPm0eq1}) and (\ref{BPm0eq2}) we obtain:

\begin{equation}\label{BPm0eq4}
    y_{\pm}(\theta)=  \frac{1}{2a}\left[1-a+2a\theta \pm  \sqrt{(1-a+2a\theta)^2-4a\theta^2}\right]
\end{equation}

\noindent where $a\equiv  {\rm tanh}(K/4)$. It is easy to see that $0\leq y_-\leq 1$ for any value of $0\leq \theta\leq 1$ and for any value of $a$. On the other hand, it can be seen that $y_+>1$ for any value of $0\leq \theta\leq 1$ when $a<0.5$. Hence, for  temperatures $k_BT/w>1/(8\, {\rm arctanh(0.5)})=0.22756$ the only meaningful solution is $y_-$.
Replacing Eqs.(\ref{BPm0eq4}) into (\ref{BPm0eq3}) we obtain two implicit equations for solving $\theta$ as a function of $h$ and $K$, namely

\begin{equation}\label{eqhm0}
    e^{-h}=F_{\pm}(\theta,a)
\end{equation}

\noindent where

\begin{equation}
    F_{\pm}(\theta,a) = 2 \left( \frac{1+y_{\pm}(\theta)-2\theta}{y_{\pm}(\theta)\,(1-a)} \right)^2 \, \left( \frac{\theta}{1-\theta} \right)^3
\end{equation}

Now, it can be seen  that $F_-(\theta,a)$ decreases monotonically with $\theta$, diverging for $\theta\to 0$ and $\lim_{\theta\to 1} F_-(\theta,a)=0$, for any value of $a$. On the other hand, $F_+(\theta,a)$ diverges both in $\theta=0$ and $\theta=1$. From the properties $y_-(0)=0$, $y_-(1)=1$, $y_+(0)=(1-a)/a$ and $y_+(1)=1/2a$, we see that the only roots of Eq.(\ref{eqhm0}) that satisfy the correct limits

\[
\theta\to 1 \;\;\;\; y\to 1 \;\;\; for  \;\;\;  h\to \infty
\]

\[
\theta\to 0 \;\;\;\; y\to 0 \;\;\; for  \;\;\;  h\to -\infty
\]

\noindent is $y_-$.

\subsection{Near the transition: susceptibilities and critical lines}

Now suppose that we add a small aligning field $B$, coupled to $m$. Now, because of the
symmetry breaking, one should consider the whole set of parallel and perpendicular
correlations, which should be different in the two principal directions of the square
lattice. Nevertheless, because of the pair approximation in (\ref{BPF}), parallel and
perpendicular correlations appear independently, i.e. the
factor involving both kinds of correlations do not mix when computing saddle point
equations. Then, the saddle point equations for each group are exactly the same, implying
that parallel and perpendicular quantities themselves are identical. Then, at least
in the Bethe-Peierls approximation, there is no symmetry breaking in the correlation
functions.

Then, from the full saddle point equations one finds:

\begin{equation}\label{BPeq7}
   3\, {\rm arctanh} \left(\frac{m}{\theta} \right) = 4\, {\rm arctanh} \left(\frac{m-z}{\theta-y} \right) - B'
\end{equation}

\noindent where $B'\equiv \beta B$. If $B'\ll 1$, we can assume $m \ll1$ and $z\ll1$ and expand

\[
3\, \frac{m}{\theta} = 4\, \frac{m-z}{\theta-y} -B' + {\cal O}(m^2,z^2,mz)
\]

\[
\frac{z}{x+y} = \frac{m-z}{\theta-y}  + {\cal O}(m^2,z^2,mz)
\]

\noindent To this order of approximation, Eqs.(\ref{BPm0eq1}), (\ref{BPm0eq2}) and (\ref{BPm0eq3}) hold. Then, we can assume $m = \chi B'$ and $z = \omega B'$. In the limit $B'\to 0$, $\chi$ is proportional to the magnetic susceptibility. Replacing in the above equations we have:

\[
3\, \frac{\chi}{\theta} = 4\, \frac{\chi-\omega}{\theta-y} -1
\]

\[
\frac{\omega}{x+y} = \frac{\chi-\omega}{\theta-y}
\]

\noindent Solving these equations we obtain

\begin{eqnarray}
  \chi &=& \frac{\theta(\theta+x)}{3x-\theta} \\
  \omega &=& \frac{\theta(x+y)}{3x-\theta}
\end{eqnarray}

\noindent where $x$ ,$y$ and $\theta$ are solutions of Eqs.(\ref{BPm0eq1})-(\ref{BPm0eq3}). The disordered solution becomes unstable whenever $3x=\theta$. Replacing these conditions into Eqs.(\ref{BPm0eq1})-(\ref{BPm0eq3}) we obtain the critical line:

\begin{equation}\label{BPcrticalh}
    e^{-h}= \frac{27}{4} \;\frac{3a-1}{1-a}
\end{equation}

\noindent Notice that, in the limit $h\to\infty$, we have $a=1/3$ or

\[
\tanh \left(\frac{K_c}{4}\right)= \frac{1}{3}
\]

\noindent which is the critical temperature for the Ising model (square lattice) in the Bethe approximation, as expected ($k_BT_c/w=0.72135$). We also have along the critical line

\begin{equation}\label{BPcrticaltheta}
    \theta_c= 3 \frac{1-a}{1+3a}
\end{equation}

\noindent or

\begin{equation}\label{BPcrticalT}
    \frac{k_BT_c}{w}= \frac{1}{4\, {\rm arctanh}\left(\frac{1}{3}\,\frac{3-\theta}{1+\theta}\right)}
\end{equation}

\section{The BP approximation in the triangular lattice case}
\label{BPtriangular}

We now consider the diluted $q=3$ anisotropic Potts model

\begin{equation}
     H = -w \sum_{<i,j>} \sum_{\sigma=1}^3 \delta(\sigma_i,\sigma)\, \delta(\sigma_j,\sigma)\, \delta(\vec{r}_{ij}, {\bf e}_\sigma) -\mu \sum_i \left[1-\delta(\sigma_i,0) \right] - \frac{B}{2} \sum_i \left[3\,\delta(\sigma_i,1)+\delta(\sigma_i,0)-1 \right]
\end{equation}

 The one site reduced matrices for these  spin variables $\sigma_i=0,1,2,3$ can be written as

\begin{equation}
    \rho_i^{(1)}(\sigma_i) = \sum_{\sigma=0}^q  P_\sigma \delta(\sigma_i,\sigma)
\end{equation}

From Eq.(37) of the manuscript we have

\begin{equation}
    \left< \delta(\sigma_i,0) \right> = P_0 = 1-\theta
\end{equation}

From Eq.(35) of the manuscript we have

\begin{equation}
    \left< \delta(\sigma_i,1) \right> = P_1 = \frac{1}{3}(\theta+2m)
\end{equation}

Using the symmetry $\left< \delta(\sigma_i,2) \right>=\left< \delta(\sigma_i,3) \right>$ and the normalization condition $\sum_{\sigma=0}^3 P_\sigma=1$ we obtain

\begin{equation}
    P_2 = P_3 = \frac{1}{2} (1-P_0-P_1)= \frac{1}{3}(\theta-m)
\end{equation}

Summarizing:

\begin{equation}
    \rho_i^{(1)}(\sigma_i) =  (1-\theta)\, \delta(\sigma_i,0)+ \frac{1}{3}(\theta+2m)\,\delta(\sigma_i,1)+\frac{1}{3}(\theta-m) \,(\delta(\sigma_i,2)+\delta(\sigma_i,3))
\end{equation}

 We next consider the two-sites reduced matrices

\begin{equation}
    \rho_{i,j}^{(2)}(\sigma_i,\sigma_j)= \sum_{\sigma,\sigma'} P_{\sigma,\sigma'} \delta(\sigma_i,\sigma)\, \delta(\sigma_j,\sigma')
\end{equation}

\noindent where we have that

\begin{equation}
    P_{\sigma,\sigma'}=\left< \delta(\sigma_i,\sigma)\, \delta(\sigma_j,\sigma')\right> = P_{\sigma',\sigma}
\end{equation}

From the reducibility conditions (\ref{reduce1})-(\ref{reduce2}) we have that

\begin{equation}
   \sum_{\sigma=0}^3 P_{\sigma,\sigma'} = P_{\sigma'} \label{psigmasigma}
\end{equation}

assuming isotropy (valid in the disordered state) we define the correlations

\begin{eqnarray}
  x &\equiv &  \left< \delta(\sigma_i,0)\, \delta(\sigma_j,0)\right> =P_{0,0}\\
  y &\equiv & \left< \delta(\sigma_i,1)\, \delta(\sigma_j,1)\right>=P_{1,1}\label{BPy1} \\
  z &\equiv& \left< \delta(\sigma_i,2)\, \delta(\sigma_j,2)\right> =\left< \delta(\sigma_i,3)\, \delta(\sigma_j,3)\right>=P_{2,2}=P_{3,3}\\
  t &\equiv& \left< \delta(\sigma_i,0)\, \delta(\sigma_j,1)\right>=P_{0,1}
\end{eqnarray}

\noindent and assuming a symmetry under interchange of states $\sigma=2$ and $\sigma=3$, and using Eqs.(\ref{psigmasigma}) we obtain

\begin{equation}\label{BPP02}
    P_{0,2}= P_{0,3}=\frac{1}{2} (P_0-x-t)= \frac{1}{2} (1-\theta-x-t)
\end{equation}

\begin{equation}
    P_{1,2}=P_{1,3}= \frac{1}{2} (P_1-y-t)= \frac{1}{2} \left[\frac{1}{3}(\theta+2m)-y-t\right]
\end{equation}

\begin{equation}
    P_{2,3}= P_2- P_{0,2}-P_{1,2}-z= \frac{2}{3}(\theta-m) +\frac{1}{2} (x+y-1)+t-z
\end{equation}

The variational BP free energy for the triangular lattice can be written as

\begin{eqnarray}
    F &=& {\rm Tr} \rho H + k_BT\left\{-5 \sum_i {\rm Tr}_i  \rho_i^{(1)} \log \rho_i^{(1)} + \sum_{<i,j>}  {\rm Tr}_{i,j}  \rho_{i,j}^{(2)} \log \rho_{i,j}^{(2)} \right\}\nonumber\\
     &=& -w N (y+2z) -\mu\theta N -B N m -5N k_BT  \sum_{\sigma=0}^3 P_\sigma \log P_\sigma + 3Nk_BT \sum_{\sigma,\sigma'} P_{\sigma,\sigma'} \log P_{\sigma,\sigma'}
\end{eqnarray}

\begin{eqnarray}\label{FBPtriangular}
    F/N &=& -w  (y+2z) -\mu\theta  -B  m - \nonumber\\
      & & -5k_BT \left\{(1-\theta) \log (1-\theta) + \frac{1}{3}(\theta+2m) \log \left[\frac{1}{3}(\theta+2m) \right] + \frac{2}{3}(\theta-m) \log \left[ \frac{1}{3}(\theta-m)\right] \right\}+ \nonumber\\
      & & + 3k_BT \left\{ x \log x + y \log y + 2 z \log z + 2t \log t + 2 (1-\theta-x-t) \log  \left[\frac{1}{2} (1-\theta-x-t)  \right]\right.  +\nonumber\\
      & & +  2\left.  \left(\frac{1}{3}\theta +\frac{2}{3}m-y-t \right) \log  \left[\frac{1}{2} \left(\frac{1}{3} \theta +\frac{2}{3} m-y-t \right) \right]+ \right.\nonumber\\
      & & +  2\left.  \left(\frac{2}{3}(\theta-m) +\frac{1}{2}(x+y-1)+t-z \right) \log  \left[\frac{2}{3}(\theta-m) +\frac{1}{2}(x+y-1)+t-z \right] \right\}
\end{eqnarray}

Deriving respect to $m$ we obtain:

\begin{equation}
    \frac{B}{2k_BT}=-\frac{5}{3} \log \left(\frac{\theta+2m}{\theta-m}\right) -2 \log \left(\frac{4(\theta-m) +3(x+y-1)+6t-6z}{\theta+2m-3(t+y)}\right)
\end{equation}

Deriving respect to $\theta$ we obtain:

\begin{eqnarray}
    \frac{\mu}{k_BT}+ \log 3&=&-\frac{5}{3} \log \left[\frac{(\theta+2m)(\theta-m)^2}{(1-\theta)^3}\right] +2 \left\{ 2 \log \left[4(\theta-m) +3(x+y-1)+6t-6z \right] + \right. \nonumber\\
    &  & + \left. \log \left[ \theta+2m
    -3(t+y)\right] -3 \log (1-\theta-x-t)\right\}
\end{eqnarray}

Deriving respect to $y$ we obtain:

\begin{eqnarray}
    \frac{w}{3k_BT}- \log 6  &=& \log y  +  \log \left[4(\theta-m) +3(x+y-1)+6t-6z \right] -  \nonumber\\
    &  & -2  \log \left[ \theta+2m -3(t+y)\right] \label{BPtriang-y1}
\end{eqnarray}

Deriving respect to $z$ we obtain:

\begin{equation}
    \frac{w}{3k_BT}-  \log 6 = \log z   -\log  \left[4(\theta-m) +3(x+y-1)+6t-6z \right] \label{BPtriang-z1}
\end{equation}

Deriving respect to $t$ we obtain:

\begin{equation}
    \log (2t) -  \log (1-\theta-x-t)  -  \log \left[ \theta+2m -3(t+y)\right]  + \log  \left[4(\theta-m) +3(x+y-1)+6t-6z \right]=0
\end{equation}

Deriving respect to $x$ we obtain:

\begin{equation}
    \log x + \log \left(\frac{2}{3}\right) -2 \log (1-\theta-x-t)    + \log  \left[4(\theta-m) +3(x+y-1)+6t-6z \right]=0
\end{equation}

Combining Eqs.(\ref{BPtriang-y1}) and (\ref{BPtriang-z1}) we find

\[
2 \log \left(\frac{4(\theta-m) +3(x+y-1)+6t-6z}{\theta+2m-3(t+y)}\right)=\log \left(\frac{z}{y} \right)
\]

\noindent and combining with the rest of equations we arrive to the following set of independent saddle-point equations:

\begin{equation}
    \frac{B}{2k_BT}=-\frac{5}{3} \log \left(\frac{\theta+2m}{\theta-m}\right) -\log \left(\frac{z}{y} \right) \label{BPSD1}
\end{equation}

\begin{eqnarray}
    \frac{\mu}{k_BT}+ \log 3&=&-\frac{5}{3} \log \left[\frac{(\theta+2m)(\theta-m)^2}{(1-\theta)^3}\right]   + \nonumber\\
    &  & + 2 \left\{ 3 \log \left[ \theta+2m -3(t+y)\right] - \log (1-\theta-x-t) -2 \log (2t)\right\} \label{BPSD2}
\end{eqnarray}

\begin{equation}
2 \log \left(\frac{4(\theta-m) +3(x+y-1)+6t-6z}{\theta+2m-3(t+y)}\right)=\log \left(\frac{z}{y} \right) \label{BPSD3}
\end{equation}

\begin{equation}
    \log x + \log \left(\frac{2}{3}\right) -2 \log (1-\theta-x-t)    + \log  \left[4(\theta-m) +3(x+y-1)+6t-6z \right]=0 \label{BPSD4}
\end{equation}

\begin{equation}
    \frac{w}{3k_BT}-  \log 6 = \log z   -\log  \left[4(\theta-m) +3(x+y-1)+6t-6z \right] \label{BPSD5}
\end{equation}

\begin{equation}\label{BPSD6}
2 \log \left(\frac{2t}{1-\theta-x-t}\right)=\log \left(\frac{y}{z} \right)
\end{equation}

\subsection{$B=0$: disordered state}

At zero field and high enough temperature we have a disordered phase, where all ordered states ($\sigma=1,2,3$) become equally probable and therefore $m=0$ ($\left<\delta(\sigma_i,1)\right>=\theta/q$). Also from the definitions (\ref{BPy1})-(\ref{BPP02}) we have that $y=z$ and $P_{0,1}=P_{0,2}$, so

\[
3t+x=1-\theta
\]

\noindent which implies that

\[
4(\theta-m) +3(x+y-1)+6t-6z= \theta+2m -3(t+y) = 2\theta +x -1 -3z
\]

With these conditions we see that Eqs.(\ref{BPSD1}), (\ref{BPSD3}) and (\ref{BPSD6}) are automatically satisfied. The remaining equations become

\begin{equation}\label{PBSDdisor1}
    \frac{\mu}{k_BT} =-5 \log \left(\frac{\theta}{3(1-\theta)}\right)   + 6 \log \left(\frac{1}{2}\,\frac{2\theta +x -1 -3z}{1-\theta-x} \right)
\end{equation}

\begin{equation}
    \frac{w}{3k_BT}- \log 6 =\log z  -  \log (2\theta +x -1 -3z)
\end{equation}

\begin{equation}
    \log x - \log \left(\frac{2}{3}\right) -2 \log (1-\theta-x)    + \log  (2\theta +x -1 -3z)=0 \label{PBSDdisor3}
\end{equation}

\begin{equation}
     t=\frac{1}{3}(1-\theta-x)\label{PBSDdisor4}
\end{equation}

\noindent which combined can be rewritten as

\begin{equation}\label{BPm0SD1}
    \frac{(1-\theta-x)^6 (1-\theta)^5}{x^6 \theta^5}= 3 e^{\beta \mu}
\end{equation}

\begin{equation}\label{BPm0SD2}
    \frac{9zx}{(1-\theta-x)^2}= e^{\beta J/3}
\end{equation}

\begin{equation}\label{BPm0SD3}
   (1-\theta-x)^2 = \frac{3}{2} x\, (2\theta+x-1-3z)
\end{equation}

\begin{equation}
     t=\frac{1}{3}(1-\theta-x)\label{BPm0SD4}
\end{equation}

From Eq.(\ref{BPm0SD1}) we can express $x$ as a function of $\theta$:

\begin{equation}
    x= \frac{(1-\theta)^{11/6}}{3^{1/6}\, e^{\beta\mu/6}\, \theta^{5/6}+ (1-\theta)^{5/6}}\label{BPm0SD4}
\end{equation}

From Eq.(\ref{BPm0SD2}) we can express $z$ as a function of $\theta$ and $x$:

\begin{equation}
    z= \frac{e^{\beta w/3}}{9x} (1-\theta-x)^2
\end{equation}

\noindent and replacing the last equation into Eq.(\ref{BPm0SD3}) we get a quadratic equation for $\theta$  in terms of $x$ whose solutions, combined with Eq.(\ref{BPm0SD4})  provide two transcendental equations for $\theta$:

\[
\theta = G_{\pm}(\theta)
\]

\noindent where

\begin{equation}\label{Gmn}
    G_{\pm}(\theta)= \frac{1}{a} \left(a+ x (3-a) \pm \sqrt{3[ax(1-x)+3x^2]}\right)
\end{equation}

\noindent where $x$ is given by Eq.(\ref{BPm0SD4}) and

\[
a \equiv 2 + e^{\beta w/3}
\]

It can be seen that the equation $\theta = G_{+}(\theta)$ has no solutions, except for $\theta=1$, where $G_{+}(\theta)=G_{-}(\theta)$. The equation $\theta = G_{-}(\theta)$ has always at least two solutions for any value of $\beta$ and $\mu$: $\theta=1$ ($x=0$) and $\theta=0$ ($x=1$). The first one is the meaningful solution in the limit $\mu\to\infty$. For large but finite values of $\mu$ a third solution with $\theta <1$ and $1-\theta \ll 1$ emerges, which decreases with $\mu$.

Let's consider the $\mu \gg1$  case. From Eq.(\ref{BPm0SD4}) we have that

\[
x \sim \frac{(1-\theta)^{11/6}}{3^{1/6}}  \, e^{-\beta\mu/6}
\]

\noindent and from Eq.(\ref{Gmn}) we have

\[
G_-(\theta) \sim 1-\sqrt{\frac{3x}{a}}
\]

Combining these results we find:

\begin{equation}\label{BPtheta-muinf1}
    1-\theta \sim \frac{e^{-\beta\mu}\, 3^5}{(2+e^{\beta w/3})^6}
\end{equation}

\begin{equation}\label{BPtheta-muinf2}
   x \sim \frac{e^{-2\beta\mu}\, 3^9}{(2+e^{\beta w/3})^{11}}
\end{equation}

\begin{equation}\label{BPtheta-muinf3}
   z \sim \frac{1}{3}\,\frac{e^{\beta w/3}}{2+e^{\beta w/3}}
\end{equation}

\subsection{Near the transition: susceptibilities and critical lines}

We now consider the case of a small external field $B' \ll 1$, where $B'\equiv \beta B$. We can assume

\[
y=z+\epsilon
\]

\[
3t+x+\theta-1=\delta
\]

\noindent where $m= {\cal O}(B')$, $\epsilon= {\cal O}(B')$ and $\delta= {\cal O}(B')$.

Then, expanding Eqs.(\ref{BPSD1})-(\ref{BPSD6}) and keeping the lowest order in $B'$, we obtain the following set of linear equations for $m$, $\epsilon$ and $\delta$

\begin{equation}\label{BPlinear1}
    \frac{6}{A} (2m-2\epsilon-\delta)= \frac{\epsilon}{z}
\end{equation}

\begin{equation}\label{BPlinear2}
    -\frac{5}{\theta} m+\frac{\epsilon}{z}= \frac{B'}{2}
\end{equation}

\begin{equation}\label{BPlinear3}
   \delta = \frac{C}{3z}\, \epsilon
\end{equation}

\noindent where

\[
A= 2\theta+x-1-3z
\]

\[
C= 1-\theta-x
\]

\noindent and $\theta$, $x$ and $z$ are the solution of Eqs.(\ref{PBSDdisor1})-(\ref{PBSDdisor3}). Solving Eqs.(\ref{BPlinear1})-(\ref{BPlinear3}), we finally obtain:

\begin{equation}\label{BPmBsmall}
    m \sim \frac{B'}{2}\; \frac{\theta (9z-x+1)}{12\theta -5(9z-x+1) }
\end{equation}

\begin{equation}\label{BPdeltaBsmall}
    \delta \sim B'\; \frac{2\theta\,(1-x-\theta)}{12\theta -5(9z-x+1) }
\end{equation}

\begin{equation}\label{BPepsilonBsmall}
    \epsilon \sim B'\; \frac{6\theta\,z}{12\theta -5(9z-x+1) }
\end{equation}

Hence, the susceptibility is

\begin{equation}\label{BPchi}
    \chi=\frac{1}{2}\; \frac{\theta (9z-x+1)}{12\theta -5(9z-x+1) }
\end{equation}

\noindent which diverges when

\begin{equation}\label{BPtriang-Tc1}
12\theta -5(9z-x+1) =0
\end{equation}

In particular, in the limit $\mu\to\infty$, when $\theta\to 1$ and $x\to 0$, the susceptibility becomes

\begin{equation}\label{BPchi}
    \chi=\frac{1}{4}\; \frac{3+4\, e^{\beta w/3}}{7-4\, e^{\beta w/3} }
\end{equation}

\noindent where we have used Eq.(\ref{BPtheta-muinf3}).